\documentclass[a4paper]{cas-sc}

\usepackage[numbers,sort]{natbib}
\usepackage{lineno,hyperref}
\modulolinenumbers[5]
\usepackage{booktabs}
\usepackage{amsmath}
\usepackage{amssymb}
\usepackage{paralist}
\usepackage{adjustbox}
\usepackage{multirow}
\usepackage{xspace}
\usepackage{color}
\usepackage{xcolor}
\usepackage{ifthen}
\usepackage{enumitem}
\usepackage{listings}
\usepackage{balance}
\usepackage{ifthen}
\usepackage{graphicx,subfigure}
\usepackage{amsmath}
\usepackage[linesnumbered,boxruled]{algorithm2e}
\usepackage{algorithm2e}
\usepackage{comment}
\usepackage{array}
\usepackage{algorithmic}
\usepackage{ulem}
\usepackage{tcolorbox}
\usepackage{amssymb} 
\usepackage{amsfonts}
\definecolor{mygray}{gray}{0.6}

\def\tsc#1{\csdef{#1}{\textsc{\lowercase{#1}}\xspace}}
\tsc{WGM}
\tsc{QE}
\tsc{EP}
\tsc{PMS}
\tsc{BEC}
\tsc{DE}

\begin{document}
\let\WriteBookmarks\relax
\def\floatpagepagefraction{1}
\def\textpagefraction{.001}
\shorttitle{Automatic Patch Linkage Detection in Code Review}
\shortauthors{D.Wang et~al.}
\title [mode = title]{Automatic Patch Linkage Detection in Code Review Using Textual Content and File Location Features}

\author[]{Dong Wang*}[type=,
                        auid=,bioid=,
                        prefix=,
                        role=,
                        orcid=]
% \ead{}
% \ead[url]{}
\credit{Conceptualization of this study, Methodology, Software}
\address[]{Nara Institute of Science and Technology, Japan}
\author{Raula Gaikovina Kula}[style=]
\credit{Data curation, Writing - Original draft preparation}
\author{Takashi Ishio}
\author%
{Kenichi Matsumoto}
% \cortext[cor1]{Corresponding author}

\newcommand{\RqOne}{(RQ1) {What is the impact of the patch linkage on the review process?}\xspace}
\newcommand{\resultBox}[1]{\begin{center}\noindent\fbox{\parbox{0.9\linewidth}{\textit{\begin{center}#1\end{ce
7
nter}}}}\end{center} }
\newcommand\wang[1]{\textcolor{red}{{\it [#1]}}}
\newcommand{\revise}[1]{\textcolor{purple}{\textbf{#1}}}
\newcommand{\RQtwo}{(RQ2) \textit{What is the performance of detecting patch linkages?}\xspace}
\newcommand{\RQthree}{(RQ2.1) \textit{What is the performance of using textual content to detect patch linkage?}\xspace}
\newcommand{\RqFour}{(RQ2.1) \textit{What is the performance of using file location to detect patch linkage?}\xspace}
\newcommand{\RqFive}{(RQ2.3) \textit{How does using more than one feature (textual content and file location) improve the performance of linkage detection?}\xspace}

\newcommand{\raula}[1]{\textcolor{blue}{{\it [#1]}}}
\newcommand{\ishio}[1]{\textcolor{magenta}{{\it [#1]}}}

\begin{abstract}
\noindent\textit{Context:} Contemporary code review tools are a popular choice for software quality assurance.
Using these tools, reviewers are able to post a \textit{linkage} between two patches during a review discussion.
Large development teams that use a review-then-commit model risk being unaware of these linkages.

\noindent\textit{Objective:} 
Our objective is to first explore how patch linkage impacts the review process.
We then propose and evaluate models that detect patch linkage based on realistic time intervals.

\noindent\textit{Method:} 
First, we carry out an exploratory study on three open source projects to conduct linkage impact analysis using 942 manually classified linkages. 
Second, we propose two techniques using textual and file location similarity to build detection models and evaluate their performance.

\noindent\textit{Results:} 
The study provides evidence of latency in the linkage notification.
We show that a patch with the Alternative Solution linkage (i.e., patches that implement similar functionality) undergoes a quicker review and avoids additional revisions after the team has been notified, compared to other linkage types.
Our detection model experiments show promising recall rates for the Alternative Solution linkage (from 32\% to 95\%), but precision has room for improvement. 

\noindent\textit{Conclusion:} 
Patch linkage detection is promising, with likely improvements if the practice of posting linkages becomes more prevalent. 
From our implications, this paper lays the groundwork for future research on how to increase patch linkage awareness to facilitate efficient reviews.
\end{abstract}


\maketitle

\section{Introduction}
Contemporary software development teams widely adopt code review tools for their software quality assurance~\cite{Bacchelli_ICSE2013,FSE2013_Rigby}.
Over the last ten years, review tools have been utilized by well-known Open Source projects such as Android, OpenStack \cite{android,openstack}, and industry giants such as Google and Microsoft~\cite{Google_2018,Bacchelli_ICSE2013}. 
In contrast to the traditional vigorous face-to-face meetings between small team members, larger teams can adopt a review tool that integrates review discussions \cite{Olga_2016,rigby_rtc_tosem}. 
Tools such as Gerrit, Codestriker, and ReviewBoard allow for patches to be submitted, and later assigned to a review team (i.e., patch author and assigned reviewers)~\cite{gerritreview,codestriker,reviewboard}.

On the downside, massive projects like OpenStack (i.e., which attracts more than 100,000 contributors that spread over 600 repositories~\cite{minghui_icse2020}) are susceptible to having their contributors submit patches that may be similar to an already existing patch.
This happens due to the parallel and distributed nature of a review-then-commit model~\cite{rigby_apache_icse}, where submitted patches achieve a similar goal (i.e., with duplication being the most typical case).
This lack of awareness occurs because review teams  (i) do not possess the knowledge of other review teams or (ii) do not notice when other authors submit similar work.

To raise awareness of similar patches during review discussions, review teams may post a \textit{linkage} between two patches to notify developers~\cite{Hirao}.
Existing work shows that the late duplicate patch identification leads to additional maintenance costs and redundant efforts~\cite{icse16_pr}.
Specifically, \citet{dataset_pull} manually studied pull requests from 26 popular projects on Github, observed that on average, 2.5 reviewers participated in the review discussions of redundant pull requests, and 5.2 review comments were generated before the duplicate relation is identified.

Apart from raising awareness of duplication, \citet{Hirao} claimed that the linkage that is posted in a review discussion can be used to point out three common patch linkage types:
\textit{Dependency} (a patch linkage to another patch that it is  dependent upon),
\textit{Broader Context} (a patch linkage to another patch that provides related resources), and 
\textit{Alternative Solution} (a patch linkage to another patch that implements similar functionality).
Their results suggest that reviewer recommendation can be improved by incorporating information from linked reviews and these linkages could be exploited by code review analytics.
Inspired by their work, we argue that the early detection of the patch linkage would help contributors to avoid unnecessary efforts, strengthen collaboration between reviewers, and encourage an effective review process.

In this paper, we extend from the idea of duplication detection to investigate the potential of detecting three different types of patch linkages (i.e., Dependency, Broader Context, and Alternative Solution).
Different from patch duplication, we do not look at the similarity of the patch source code, but instead focus on linkages between two patches.
Through a case study of three large Open Source projects, i.e., Qt, OpenStack, and Android Open Source Project (AOSP), 
we are able to extract 11,353 patch linkages in total.
We formulate two research questions to guide our study:
\begin{itemize}
    \item \textbf{\RqOne}\\
    \textit{\uline{Motivation.}}
    \citet{Hirao} investigated the different types of linkage by classifying their various purposes in the review discussion.  
    However, the impact of the linkage on the review process is unknown.
    Specifically, we would like to investigate the impact in terms of when the linkage is first notified (i.e., submission time to notification time) and the time taken for that review (i.e., notification time to decision time).
    Moreover, we would like to investigate whether or not the entire review process (i.e., submission time to decision time) of patches with patch linkages is different from those patches with no patch linkages.
    To address this RQ, we conduct an exploratory study.
    \item \textbf{\RQtwo}\\
    \textit{\uline{Motivation.}}
    Motivated by the exploratory study (RQ1),
    we would like to evaluate whether or not it is promising to detect these patch linkages.
    Although recent work~\cite{pull_request17, ren_saner2019, xinxia_pull} investigate the feasibility of detecting pull request in the code review setting, it is still unclear that how linkage detection performs in a realistic setting, other than duplication.
    We assume that patches that are linked tend to share similar textual content and modify similar code locations.
    Thus, we use these two patch features to construct our detection model borrowed from the information retrieval-based and file location-based recommendation.
    Taking realistic evaluation into account, we evaluate our models based on four time intervals (i.e., 2, 7, 14, and 30 days).
    This RQ is split into three sub-questions:
    \begin{itemize}
    \item \textbf{\RQthree}
    \item \textbf{\RqFour}
    \item \textbf{\RqFive}
    \end{itemize}
\end{itemize}

The key results of each RQ are: For RQ1, 
results show that there exists latency in the notification of linked patches (i.e., a median of 1.2 days, 3.1 days, and 2.2 days for Qt, OpenStack, and AOSP).
Results also show that patches with linkages are likely to take a longer time to review when compared to a control group (i.e., patches without linkages), as patches without linkages seem to not require additional reviewing efforts (23 out of 28 patches).
Earlier patch linkage notification could make for a more efficient review, especially in detecting the Alternative Solution.
For RQ2, results show that combining two features (i.e., textual content and file location) performs better than two separate models.
In experiments that span four time intervals, the model performs with promising recall rates (i.e., 24\%--68\% for Qt, 25\%--61\% for OpenStack, and 28\%--81\% for AOSP).
The Alternative Solution linkage detection is also promising with relatively high recall rates (i.e., 74\%--95\% for Qt, 71\%--87\% for OpenStack, and 77\%--94\% for AOSP in the Top-10).
Reasonable Alternative Solution linkage detection also means that the precision rates are feasible,  with 60\%--74\% for Qt, 43\%--67\% for OpenStack in the textual content model, and  56\%--67\% for AOSP in the file location model.

We suggest that improving the awareness between the patches may also increase the likelihood that the linked patches will be identified.
To improve the textual content model, developers could be encouraged to increase the natural language or generate a project specific corpus.
We also see that linkage detection is promising in a realistic setting, especially for the Alternative Solution linkage. 
The main contributions of this paper are three-fold:
First, our exploratory study provides evidence that the latency of the linkage notification exists.
Second, we propose and evaluate our detection models for three different types of patch linkages.
Finally, we provide a replication package which includes (a) manually coded patch linkages with various properties and (b) experiment datasets and scripts that can be used to reproduce our detection model.

\begin{figure}[pos = t]
    \centering
    \subfigure[Submitted patch with textual content and file location. \#86771]{\includegraphics[width=0.5\linewidth]{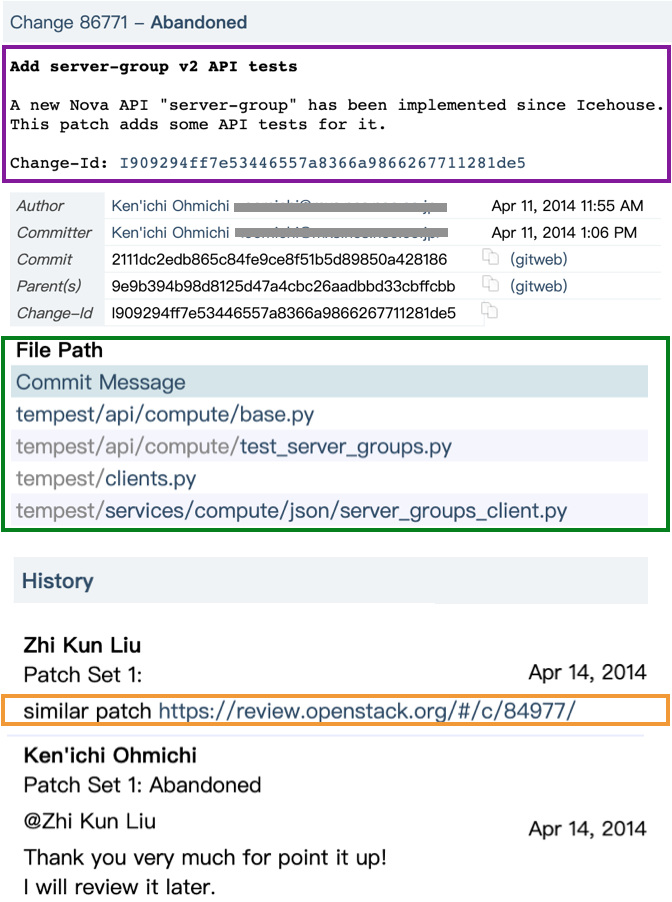}}
    \subfigure[Target patch with textual content and file location. \#84977]{\includegraphics[width=0.5\linewidth]{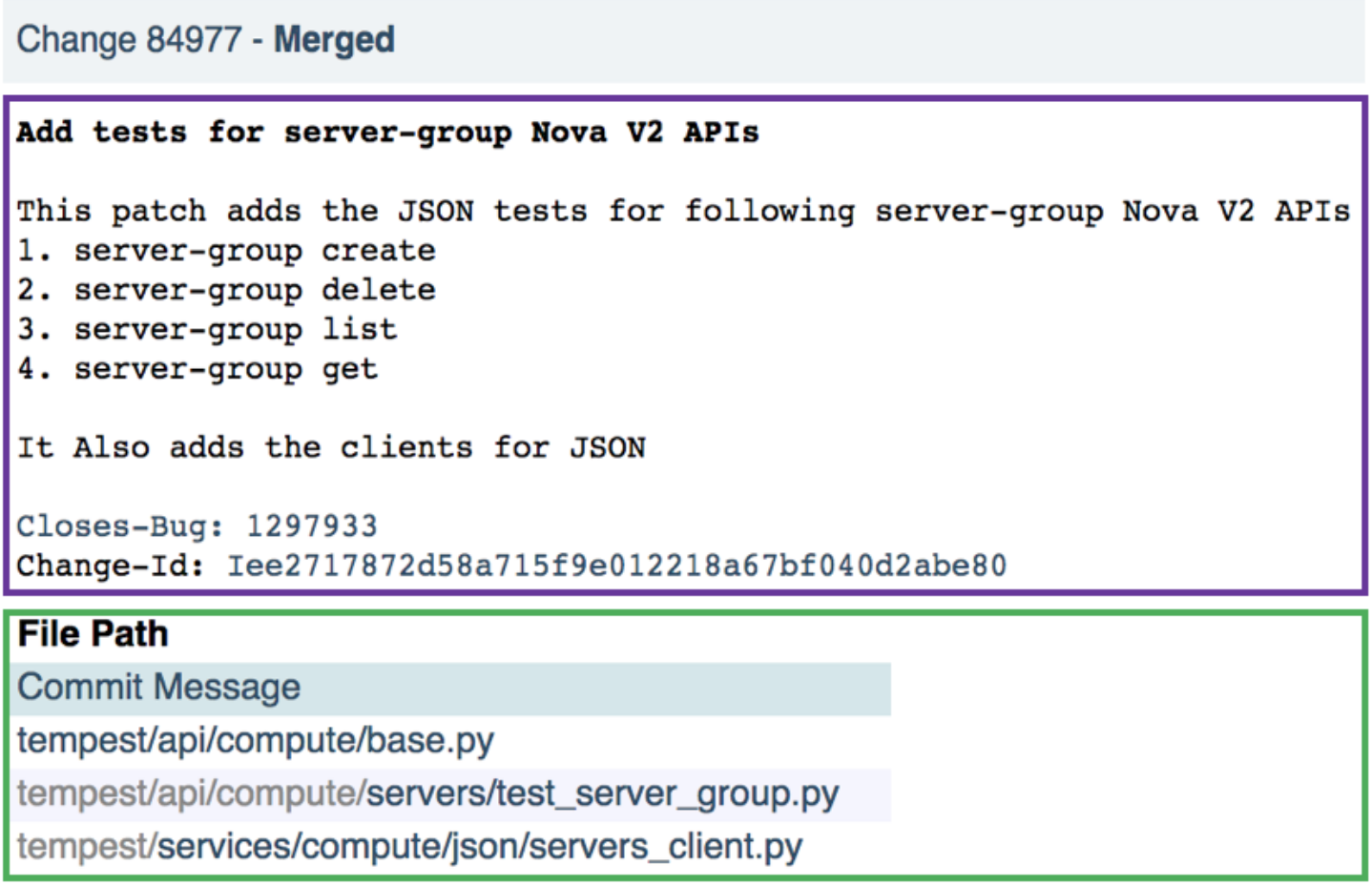}}
    \caption{A real world example to motivate the Alternative Solution linkage between patch \#86771 and patch \#84977 in OpenStack. The example suggests that the linked patches share similar textual content and modify similar set of file paths.}
    \label{fig:example}
\end{figure}

The remainder of the paper is structured as follows:
Section \ref{sec:Motivating_example} describes the motivating example. 
Section \ref{sec:exploratory_study} presents an exploratory study on the impact of the patch linkage on the review process.
Section \ref{sec:linkagedetection} introduces the patch linkage detection.
Section \ref{sec:implications} discusses the implications from our findings.
Section \ref{sec:threats} discloses threats to the validity of our study.
Related work is discussed in Section \ref{sec:relatework}.
Finally, we draw our conclusions in Section \ref{sec:conclusion}.

\section{Motivating Example}
\label{sec:Motivating_example}
Figure \ref{fig:example} is a real-world example to illustrate how a reviewer posts a linkage to notify the review team there is a similar patch.
In the figure, one author (Ken'ichi Ohmichi) submitted a patch \# 86771 to the review tool. 
The patch \# 86771 aims to add API tests in a new Nova API called ``server-group''.
During a review discussion, the reviewer (Zhi Kun Liu) pointed out that this patch addressed an issue similar to another patch:  \textit{`similar patch \url{https://review.openstack.org/\#/c/84977/}'}
and provided an explicit link of that patch.
Once notified, the author (Ken'ichi Ohmichi) proceeded to abandon the patch without any revisions on that same day.
From the description of the patch \#84977, we observe that its goal is as well to add several tests for ``server-group'' Nova APIs.
From the activity log, three days passed before the reviewer identified and posted the linkage into the patch \# 86771.
The example provides evidence that latency exists before the team is notified to become aware of the linkage, i.e., around three days are taken in the motivating example.

In terms of the feature similarity between two linked patches, we find that the linked patches from the example share similar textual content and modify similar file locations.
As shown in the figure, the two patches use the same keywords (i.e., \texttt{server-group}, \texttt{Nova}, \texttt{v2}, \texttt{tests}).
Apart from highly similar textual content, both patches touch the same file path (i.e., tempest/api/compute/base.py).

\section{Impact of Patch Linkage on the Review Process: An Exploratory Study}
\label{sec:exploratory_study}
To answer RQ1, we conduct an exploratory study to investigate the impact of patch linkage on the review process.
In this section, we first describe the studied projects (Section 3.1), then we describe the data preparation (Section 3.2),  and present the analysis approach (Section 3.3).
Finally, we discuss our results (Section 3.4).

\begin{table}[pos = b]
\caption{Collected dataset including three open source projects: Qt, OpenStack, and AOSP. In total, 11,353 patch linkages are retrieved from these projects.}
\centering
\begin{tabular}{l|r|r|r}
\hline
\textbf{}              & \textbf{Qt} & \textbf{OpenStack} & \textbf{AOSP} \\
Time Period                             & Aug.2011 $\sim$ Apr.2015     & Aug.2011 $\sim$ Nov.2016            & Oct.2008 $\sim$ Apr.2015       \\ \hline
\# Sub-Projects                         & 111                          & 1,528                               & 567                            \\
\# Files                                & 176,898                      & 158,168                             & 166,931                        \\
\# Patches                              & 107,858                      & 215,725                             & 59,490                         \\
\# Revisions                            & 258,066                      & 766,017                             & 113,501                        \\
\# Lines of Code (LOC) $\triangle$      & 273,395                      & 8,753,713                           & 35,202,362                     \\ \hline
\# Patch linkages (Hyperlinks)          & 705                          & 8,048                               & 1,079                          \\
\# Patch linkages (Review \#, Changeid) & 253                          & 1,002                               & 266                            \\
Total                                   & (0.8\%) 958                  & (4.2\%) 9,050                       & (2.2\%)  1,345                 \\ \hline
\end{tabular}
\label{datasets}
\end{table}

\subsection{Studied Projects}
We select studied projects with the two criteria: (1) large software projects actively use review tools (e.g., Gerrit) and  (2) projects are representative and have been commonly analyzed in prior work.
From the range of open source projects, we select three projects: Qt, OpenStack, and Android Open Source Project (AOSP) as these three projects actively perform code reviews through Gerrit.
Qt is a cross-platform application for creating graphical user interfaces.
OpenStack is a collaborative platform for cloud computing, which is used by many well-known organizations and companies (e.g., IBM, VMware, and NEC).
Finally, AOSP is a mobile operating system developed by Google.

\subsection{Data Preparation}

Our data  preparation process consists of three parts: (DP1) patch linkage recovery and filtering, (DP2) ground-truth construction, and (DP3) control group construction.
We define the patch linkage as any unique linkage from a patch to another. 
Let two patches be $\mathbb{p}_a$ and $\mathbb{p}_b$, where
$\mathbb{p}_a$ refers to a patch where the review team posts a linkage into the review discussion and $\mathbb{p}_b$ denotes the target patch.
Therefore, a linkage from $\mathbb{p}_a$ to target $\mathbb{p}_b$ and a linkage from $\mathbb{p}_b$ to target $\mathbb{p}_a$ are counted separately.

\begin{itemize}
\item {\uline{\textit{(DP1) Patch linkage recovery and filtering:}}}
For studied projects, we adopted the \citet{MSR2016_Yang} dataset as our original dataset.
Table \ref{datasets} shows our dataset summary (i.e., 107,858 patches for Qt, 215,725 patches for OpenStack, and 59,490 patches for AOSP).
To recover the patch linkage, we make sure that all linkage formats should be considered including Changeid, Review \# as well as the hyperlinks.
For the hyperlink format, we use regular expression patterns to search all discussions (e.g., \textit{https://review.openstack.org/\#/c/84977/}).
Additionally, we conduct manual checks on each review carefully to ensure the number is correctly related to the Changeid and Review \#, but not for other information such as bug id. 
For example, a detected Changeid would be \textit{ "would conflict with 9afb02412eadc567e82a0aca10c6401937d213e9"}, while an example of a detected Review \# is ``\textit{Replaced by 22724, 22725}".

Furthermore, we apply three filters to ensure an unbiased dataset.
The first filter is to remove cases where the patch author is aware of the linkage.
This is done by using two conditions.
The first condition is to detect the case where linked patches are written by the same author.
The second condition is to detect the case where a patch author cherry-picks or reverts an already known patch.
The second filter is to remove patches with incomplete author information (i.e., where the authors could not be retrieved from the REST API).
The third filter involves the removal of duplicate patch linkages.
We only detect the first instance of a linkage, as the same patch linkage can be used several times by different reviewers during review discussions.
After DP1, we are left with 1,345, 9,050, and 958 distinct linkages posted by different authors as shown in Table \ref{datasets}.

\item {\uline{\textit{(DP2) Ground-truth construction:}}} 
To construct our ground-truth, we manually classify the three different types of patch linkages (i.e., Alternative Solution, Broader Context, and Dependency) using the linkage taxonomy defined by \citet{Hirao}.
Since our collected data from DP1 is too large to manually examine, we statistically generate a representative sample using a statistical calculator~\cite{stas_cal} with a 95\% confidence level and a margin error of no more than 5\%, which is similar to previous empirical studies in the SE domain~\cite{Hata:ICSE19,news_aggregators}.
We end up with 369 patch linkages, 299 patch linkages, and 274 patch linkages as shown in Table \ref{tab:Representative2}.

In the coding process, we first test our comprehension with a statistical agreement.
Three authors of this article independently coded a random sample of 30 comments that contain the patch linkages based on the constructed coding scheme.
We then measured the Kappa agreement for the coding results.
The Kappa statistic is used frequently to measure inter-rater reliability for qualitative (categorical) items~\cite{viera2005understanding}.
We ended up with a Kappa score of 0.83, which indicates that our agreement is nearly perfect. 
Two authors then completed the coding for the remaining sample dataset.
We classify those patch linkages which do not indicate Broader Context, Dependency, and Alternative Solution into Others Category.
After DP2, 317 patches, 311 patches, and 233 patches are labeled as either Alternative Solution, Broader Context, and Dependency.

\item {\uline{\textit{(DP3) Control group construction:}}} To construct a balanced control group, we then randomly select an equal 861 patches from the ground-truth in DP2, as shown in Table 2.
We carefully and manually checked each patch to ensure that there was no indication of a linkage to another patch.

\begin{table}[pos = b]
\footnotesize
\caption{Ground-Truth based~on \citet{Hirao} and Control Group (patches with no patch linkages).}
\centering
\begin{tabular}{lrrrrrr}
\cline{2-7}
\multicolumn{1}{l|}{}           & \multicolumn{5}{c|}{Ground-Truth}                                                               & \multicolumn{1}{l|}{Control Group} \\ \hline
\multicolumn{1}{|l|}{Project}   & Alternative Solution & Broader Context & Dependency & Others & \multicolumn{1}{r|}{Sample Size} & \multicolumn{1}{r|}{Non-PatchLink} \\
\multicolumn{1}{|l|}{Qt}        & 93                   & 77              & 74         & 30     & \multicolumn{1}{r|}{274}         & \multicolumn{1}{r|}{244}           \\
\multicolumn{1}{|l|}{OpenStack} & 97                   & 147             & 101        & 24     & \multicolumn{1}{r|}{369}         & \multicolumn{1}{r|}{345}           \\
\multicolumn{1}{|l|}{AOSP}      & 127                  & 87              & 58         & 27     & \multicolumn{1}{r|}{299}         & \multicolumn{1}{r|}{272}           \\\hline
Total                           & 317                  & 311             & 233        & 81     & 942                              & 861                               
\end{tabular}
\label{tab:Representative2}
\end{table}
\end{itemize}

\subsection{Analysis for RQ1}
To analyze the impact of patch linkage on the review process, we focus on two main aspects: (1) comparison among linkage types and (2) comparison against a control group (patches with no patch linkages).
Below, we describe our analysis approach for each aspect.

\uline{\textit{Comparison among linkage types}.} 
We now investigate the three different patch linkage types, in terms of their review process (i.e., review time, patch revisions before and after the patch linkage notification).
We define four metrics to conduct our statistic analysis as shown below: 
\begin{itemize}
    \item \textit{First-Notify-Time (\# days)} is the duration from when the author submits $\mathbb{p}_a$ until when the link to $\mathbb{p}_b$ first appears in the review discussion.
    \item \textit{First-Notify-Revisions}  is the number of patch revisions that occur after the author submits a $\mathbb{p}_a$ until the link to $\mathbb{p}_b$ first appears in the review discussion.
    Note that the linkage is always mapped to the current patch revision, thus the first submission is counted as the first patch revision.
    \item \textit{Notify-to-Decision-Time (\# days)} is the duration from when the link of $\mathbb{p}_b$ first appears until there is a decision to either abandon or merge $\mathbb{p}_a$ into the codebase.
    \item \textit{Notify-to-Decision-Revisions} is the number of patch revisions that occur after the link to $\mathbb{p}_b$ first appears until there is a decision to either abandon or merge $\mathbb{p}_a$ into the codebase.
\end{itemize}
After the metric computation, we then test the hypothesis `\textit{a patch with the Alternative Solution linkage is reviewed quicker after notification when compared with other two linkages.'}
We use a Mann-Whitney U test ($\alpha$= 0.05)~\cite{mann1947} to validate our hypothesis and investigate the effect size using Cliff’s Delta~\cite{Cliff1996}.
Effect size is analyzed as follows: (1) $|\delta|$~\textless~0.147 as Negligible, (2) 0.147~$\le$~$|\delta|$~\textless 0.33 as Small, (3) 0.33~$\le$ $|\delta|$~\textless 0.474 as Medium, or (4) 0.474~$\leq$~$|\delta|$~as Large.

\uline{\textit{Comparison against control group (patches with no patch linkages)}.}
In this aspect, we compare the review process between patches with linkages and patches that do not have any patch linkages with two additional metrics.
Two metrics related to the entire review process are defined as follows:
\begin{itemize}
    \item \textit{Submit-to-Decision-Time (\# days)} is the duration from  when the author submits a patch until there is a decision to either abandon or merge the patch into the codebase.
    \item \textit{Submit-to-Decision-Revisions} is the number of patch revisions from when the author submits a patch until there is a decision to either abandon or merge the patch into the codebase.
\end{itemize}
After the metric computation, we test our hypothesis that `\textit{there exists a difference between patches with linkages and those that do not have any patch linkages with regard to the reviewing time}'.
We use Kruskal-Wallis non-parametric statistical test~\cite{Kruskal-Wallis} to validate our hypothesis.

\subsection{Results for RQ1}
We analyze the impact of the patch linkage on the review process in terms of the comparison among patch linkage types and the comparison against the control group (patches with no patch linkages).
Table 3 and Figure 3 show the related results.
We now discuss our results below.

\begin{figure}[pos = t]
    \centering
    \subfigure[ \textit{Notify-to-Decision-Time} Distribution.
    \label{RQ1_plot_a}
    ]{\includegraphics[width=.48\linewidth]{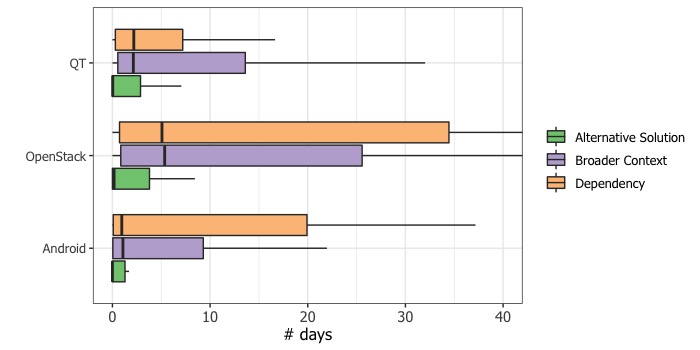}}
    \subfigure[ \textit{Notify-to-Decision-Revisions} Distribution.
    ]{\label{RQ1_plot_b}\includegraphics[width=.48\linewidth]{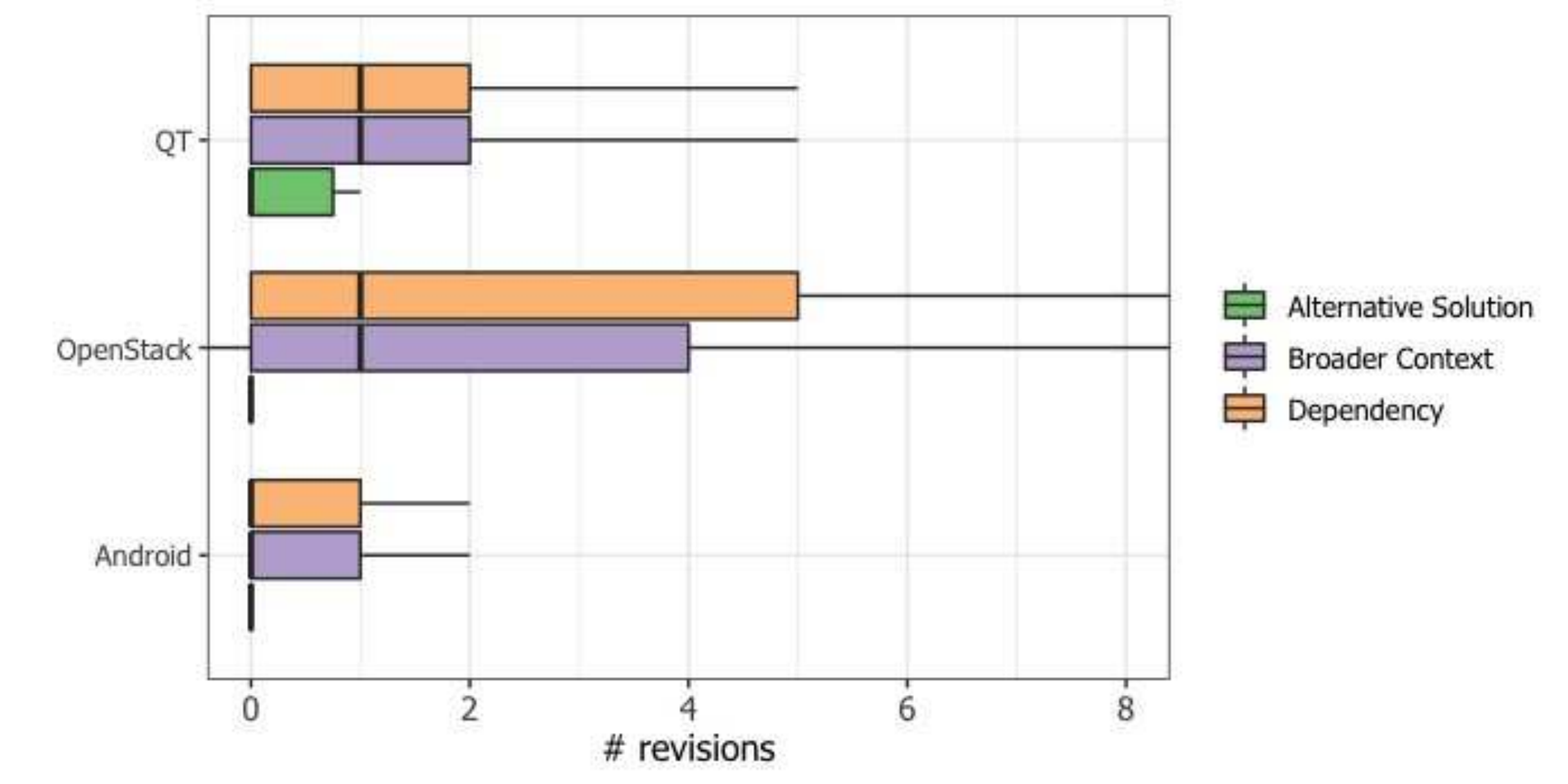}}
    \caption{Box-plots showing comparison among linkage types (\textit{Notify-to-Decision-Time} and \textit{Notify-to-Decision-Revisions}). The results show that the patch with an Alternative Solution linkage tends to have a quicker review process after the notification, compared with other patch linkages.}
    \label{RQ1_plot}
    \subfigure[\textit{Submit-to-Decision-Time} Distribution.
    \label{RQ2_plot_a}
    ]{\includegraphics[width=.5\linewidth]{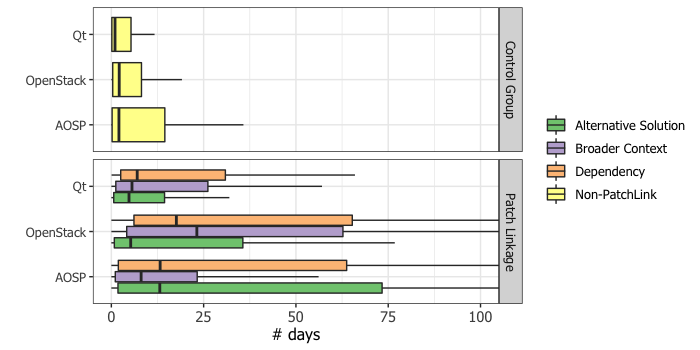}}
    \subfigure[\textit{Submit-to-Decision-Revisions} Distribution.
    ]{\label{RQ2_plot_b}\includegraphics[width=.48\linewidth]{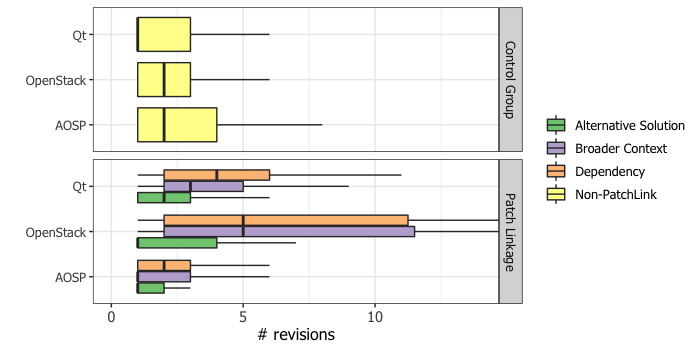}}
    \caption{Box-plots showing comparison against a control group (\textit{Submit-to-Decision-Time} and \textit{Submit-to-Decision-Revisions}).The results show that compared to patches having no linkages, patches with linkages tend to take a longer time to complete the review process.
    }
    \label{RQ2_plot}
\end{figure}

\begin{table}[pos = hbt!]
\footnotesize
\centering
\caption{Statistics showing comparison among linkage types (\textit{First-Notify-Revisions} and \textit{First-Notify-Time}). The results suggest that latency exists in the notification of a patch linkage (i.e.,  the median of 1.2, 3.1, and 2.2 days for Qt, OpenStack, and AOSP).
}
\begin{tabular}{ll|rrr|rrr|}
\cline{3-8}
                                                &                      & \multicolumn{3}{c|}{First-Notify-Revisions } & \multicolumn{3}{c|}{First-Notify-Time (\# days)} \\ \hline
\multicolumn{1}{|c}{Project}                    & Linkage Type         & Median            & Mean           & Max           & Median            & Mean           & Max             \\ \hline
\multicolumn{1}{|c}{\multirow{4}{*}{Qt}}        & Alternative Solution & 1                 & 2              & 18            & 1.1               & 16.4           & 261.6           \\
\multicolumn{1}{|c}{}                           & Broader Context      & 2                 & 3              & 21            & 1.1               & 21.4           & 627.8           \\
\multicolumn{1}{|c}{}                           & Dependency           & 2                 & 5              & 56            & 2.6               & 20.4           & 400.5           \\
\multicolumn{1}{|c}{}                           & All                  & 2                 & 3              & 56            & 1.2               & 19.2           & 627.8           \\ \hline
\multicolumn{1}{|c}{\multirow{4}{*}{OpenStack}} & Alternative Solution & 1                 & 3              & 29            & 1.1               & 26.5           & 448.0           \\
\multicolumn{1}{|c}{}                           & Broader Context      & 2                 & 5              & 28            & 3.5               & 22.6           & 418.0           \\
\multicolumn{1}{|c}{}                           & Dependency           & 3                 & 6              & 43            & 4.3               & 26.8           & 574.8           \\
\multicolumn{1}{|c}{}                           & All                  & 2                 & 4              & 43            & 3.1               & 25.0           & 574.8           \\ \hline
\multicolumn{1}{|c}{\multirow{4}{*}{AOSP}}   & Alternative Solution & 1                 & 2              & 8             & 6.2               & 48.7           & 669.0           \\
\multicolumn{1}{|c}{}                           & Broader Context      & 1                 & 2              & 9             & 0.6               & 13.2           & 258.7           \\
\multicolumn{1}{|c}{}                           & Dependency           & 1                 & 2              & 14            & 0.8               & 32.3           & 771.0           \\
\multicolumn{1}{|c}{}                           & All                  & 1                 & 2              & 14            & 2.2               & 33.5           & 771.0           \\ \hline
\end{tabular}
\label{RQ1}
\end{table}

\uline{\textit{Comparison among linkage types}.} Latency exists in the notification of a patch linkage.
Table \ref{RQ1} shows the statistics of \textit{First-Notify-Time} and \textit{First-Notify-Revisions}.
Using \textit{First-Notify-Time} metric,  we find that it takes a median of 1.2, 3.1, and 2.2 days for the linkage to be notified for Qt, OpenStack, and AOSP.
Comparing the projects, we find that OpenStack tends to have more linkage latency compared with the other two projects.
We find that the time latency is up to around 3.5 and 4.3 days (i.e., median) for Broader Context and Dependency linkages, while in AOSP, the Alternative Solution linkage takes almost 6.2 days before the patch is made known to the review team. 
On the other hand, using the \textit{First-Notify-Revisions} metric, we observe that the author already revised the patch before a patch linkage was notified, especially for Qt and OpenStack.

The patch with an Alternative Solution linkage tends to have a quicker process after the notification.
Figure \ref{RQ1_plot} shows the comparison from the time when the review team is notified until the decision for the patch has been made.
Figure \ref{RQ1_plot_a} shows evidence that a patch with the Alternative Solution linkage tends to be reviewed quicker when compared to the other patch linkages.
For a patch with the Broader Context or Dependency linkage, it takes a longer time to complete the review.
Based on our experience with manual coding,
a potential reason is due to the nature of both patch linkages.
Furthermore, Figure \ref{RQ1_plot_b} shows that a patch with the Alternative Solution linkage will be resolved without patch revisions after the notification.
For instance, the median of \sloppy{\textit{Notify-to-Decision-Revisions}} for the Alternative Solution linkage is 0 for all projects, while for Broader Context and Dependency linkages, patches with them tend to have additional revisions, i.e., one more patch revision in Qt and OpenStack.
For the statistical test, Mann-Whitney U tests confirm that `\textit{a patch with the Alternative Solution linkage is quicker to complete the review after the notification compared with other two linkages'} with \textit{p}-value~\textless~0.001 for the three studied projects.
Additionally, the Cliff’s Delta scores indeed prove our hypothesis, i.e., $|\delta|$=0.43 (medium) for Qt, $|\delta|$=0.48 (large) for OpenStack,  and $|\delta|$=0.36 (medium) for AOSP.

\uline{\textit{Comparison against control group (patches with no patch linkages)}.}
Compared to those patches with no linkages, patches with linkages tend to take a relatively longer time to complete the review process.
Figure \ref{RQ2_plot_b} presents the distribution of \textit{Submit-to-Decision-Revisions}. The results show that a patch with the Broader Context or Dependency linkage is likely to have more patch revisions than our control group.
On the other hand, Figure \ref{RQ2_plot_a} shows that patches with linkages tend to have longer reviewing time compared to those that do not have any patch linkages.
It may mean that patches including linkages require more review discussion~\cite{Hirao_tse,saner_confusion2019}, although our results suggest that the Alternative Solution linkage does not require as much discussion or revisions. 
For the statistical test, the values of Kruskal-Wallis tests reveal that `\textit{there exists a difference between patches with linkages and those that do not have any patch linkages with regard to the reviewing time}'.

Inspired by the results, we would like to further explore why the review process of the patch with no linkages is quicker than the patch with an Alternative Solution linkage.
To do so, we randomly select 30 samples (i.e., the median of \textit{Submit-to-Decision-Time} is smaller than the median time for the Alternative Solution linkage), covering three studied projects from the control group.
Then we conduct a qualitative analysis to investigate these 30 samples in the aspect of the reviewer participation~\cite{EMSE2017_Pick}, review divergence~\cite{Hirao_tse}, and review confusion~\cite{saner_confusion2019}.
The analysis results show that within these 28 patches (ignoring 2 self- approved patches), the reviewer response is timely while few comments are left (i.e., 20 out of 28 patches have less than 3 comments), review divergence rarely exists (i.e., all 28 patches do not have the case where reviewers have divergence), and the confused reviews seldom occur (i.e., 5 out of 28 patches have confusion).
The qualitative findings suggest that the review process of the patch with no linkages is relatively straightforward, which could explain why the patch with no linkages is likely to take a shorter time compared to the patch with linkages.

\begin{tcolorbox}
\textbf{Answering RQ1:} Our exploratory results show latency in the notification of linked patches (i.e., a median of 1.2 days, 3.1 days, and 2.2 days for Qt, OpenStack, and AOSP).
Results also show that patches with linkages are likely to take a longer time to review when compared to a control group (i.e., patches without linkages), as patches without linkages seem to not require additional reviewing efforts (23 out of 28 patches).
Earlier patch linkage notification could make for a more efficient review, especially in detecting the Alternative Solution.
\end{tcolorbox}

\begin{figure}[pos = t]
    \centering
    \includegraphics[width=0.7\linewidth]{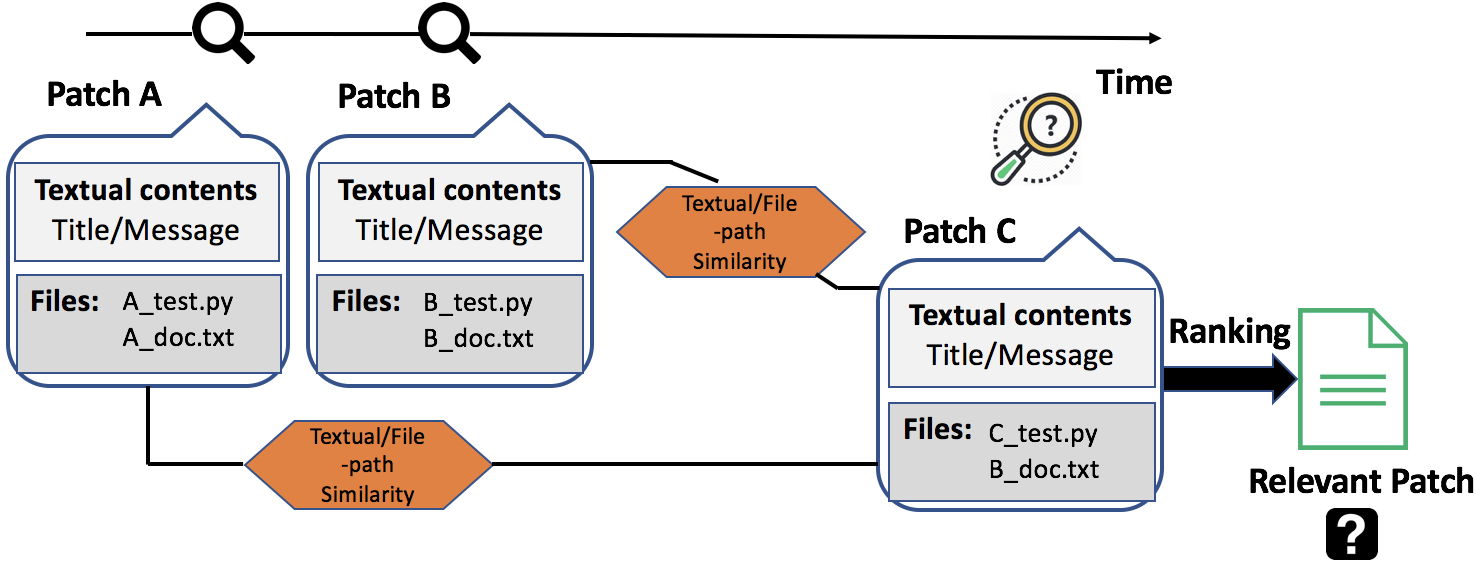}
    \caption{The overview of our linkage detection process. To calculate the similarity between the two patches, we focus on the following features: textual content feature (the concatenation of title and description text in a patch) and file location feature (a set of file paths that the patch modifies).}
    \label{fig:overall}
\end{figure}

\section{Patch Linkage Detection}
\label{sec:linkagedetection}
To answer RQ2, we propose techniques using two patch features (i.e., textual content and file location) to detect the patch linkage.
Particularly, we aim to detect the Alternative Solution linkage, which has potential for an efficient review.
In this section, we first provide a technique overview (Section 4.1), then introduce our experiment dataset (Section 4.2), and describe the evaluation metrics (Section 4.3).
Finally, we discuss our results (Section 4.4).

\subsection{Technique Overview} 
Figure \ref{fig:overall} presents an overview of our proposed process to detect patch linkage, focusing on the patch itself.
Our key assumption is that linked patches share similar textual contents and modify similar file locations.
We define the textual content as the concatenation of title and description text in a patch. 
The file location refers to a set of file paths that the patch modifies.
For instance, in the figure, the file location of the Patch A is [A\_test.py, A\_doc.py].

\paragraph{Detection Using Textual Content - }
The first technique is to compute the textual content similarity.
Since the length of the textual content is relatively short, it is appropriate to use the vector space model (VSM), which works well and is widely applied in a similar context such as duplicate bug reports~\cite{Thung_2014}.
The first step involves text processing, which includes tokenization, stop word removal, and stemming.
Next, in the representation step, our pre-processed text is converted as a vector weight representation.
Then we employ a term
frequency-inverse document frequency (\textit{tf-idf)}~\cite{Ramos2003UsingTT} as weighting scheme and calculate the cosine similarity~\cite{Manning_2008} between two linked patches.

\begin{algorithm}[hbt!]
\footnotesize
    \SetKwInOut{Input}{Input}
    \SetKwInOut{Output}{Output}
    \SetKwInOut{Method}{Method}
    \Input{\noindent A new patch ($\mathbb{p}_n$); A set of patches ($Patches$)}
    \Output{A list of candidate patches}
    $\mathbb{T}_n \leftarrow Tokenize(\mathbb{p}_n.textual content)$\;
    Remove stop words from $\mathbb{T}_n$\;
    Stem each word in $\mathbb{T}_n$\;
    $\mathbb{v}_n \leftarrow ConstructVSM(\mathbb{T}_n)$\;
    \For{$\mathbb{p}_i \in Patches$}{
    $\mathbb{T}_i \leftarrow Tokenize(\mathbb{p}_i.title, \mathbb{p}_i.message)$\;
    Remove stop words from $\mathbb{T}_i$\;
    Stem each word in $\mathbb{T}_i$\;
    $\mathbb{v}_i \leftarrow ConstructVSM(\mathbb{T}_i)$\;
    $\mathbb{sim}_i \leftarrow CosineSimilarity(\mathbb{v_n}, \mathbb{v}_i)$\; 
    }
    \Return A patch list in $Patches$ in the descending order of $\mathbb{sim}_i$;
  \caption{Detect linkage based on textual content similarity}
  \label{Al1}
\end{algorithm}
%%%%%%%%%%%%%%%%%%%%%%%%%%%%%%%%%%%%%

Based on \citet{Runeson07_ICSE}, Algorithm \ref{Al1} details our procedure to rank patches based on textual content similarity.
Inputs are new patch $\mathbb{p}_n$ and a set of patches $Patches$.
In detail, we first perform the text processing for $\mathbb{p}_n$ (i.e., the concatenation of title and description text)
using NLTK\footnote{\url{https://www.nltk.org/}} package and well known Porter stemming~\cite{Porter} (lines 1-3).
Then we apply gensim\footnote{\url{https://radimrehurek.com/gensim/}} package to create a VSM representation defined as $\mathbb{v}_n$ weighted by \textit{tf-idf} (line 4).
To process the textual content for each $\mathbb{p}_i$ $\in$ $Patches$, we construct a VSM representation defined as $\mathbb{v}_i$ (lines 5-9).
We compute the similarity between $\mathbb{v}_n$ and $\mathbb{v}_i$ $\in$ $Patches$ through \textit{cosine similarity} (line 10) and then sort the patches in $Patches$ based on their cosine scores (line 12), returning a resultant list of candidate patches in descending similarity order.

\paragraph{Detection Using File Location - }
The second technique is to compute the file location similarity.
We apply the file location-based model which performs well in reviewer recommendation task~\cite{SANER2015_Pick, Yu_2016_reviewer}.
\begin{table}[pos = hbt!]
\footnotesize
\caption{File path comparison technique descriptions. Similar to the work of ~\citet{SANER2015_Pick} four comparison techniques are included: LCP, LCS, LCSubstr, and LCSubseq.}
\begin{tabular}{|l|l|l|}
\hline
Functions                                                                         & Description                                                                                                                                         & Example                                                                                                                                                                                                  \\ \hline
\begin{tabular}[c]{@{}l@{}}Longest Common Prefix\\ (LCP)\end{tabular}             & \begin{tabular}[c]{@{}l@{}}Longest consecutive path components\\ that appears in the beginning {of} both\\ file paths.\end{tabular}                   & \begin{tabular}[c]{@{}l@{}}f1 =“src/com/android/settings/LocationSettings.java”\\ f2 = “src/com/android/settings/Utils.java”\\ LCP(f1, f2) = length({[}src, com, android, settings{]}) = 4\end{tabular}                                                                \\ \hline
\begin{tabular}[c]{@{}l@{}}Longest Common Suffix\\ (LCS)\end{tabular}             & \begin{tabular}[c]{@{}l@{}}Longest consecutive path components\\ that appears in the end {of} both file\\ paths\end{tabular}                          & \begin{tabular}[c]{@{}l@{}}f1 = “src/imports/undo/undo.pro”\\ f2 = “tests/auto/undo/undo.pro”\\ LCS(f1, f2) = length({[}undo, undo.pro{]}) = 2\end{tabular}                                                                                                            \\ \hline
\begin{tabular}[c]{@{}l@{}}Longest Common\\ Substring (LCSubstr)\end{tabular}     & \begin{tabular}[c]{@{}l@{}}Longest consecutive path components\\ that appears in both file paths\end{tabular}                                       & \begin{tabular}[c]{@{}l@{}}f1 = “res/layout/bluetooth\_pin\_entry.xml”\\ f2 = “tests/res/layout/operator\_main.xml”\\ LCSubstr(f1, f2) = length({[}res, layout{]}) = 2\end{tabular}                                                                                    \\ \hline
\begin{tabular}[c]{@{}l@{}}Longest Common\\ Subsequence\\ (LCSubseq)\end{tabular} & \begin{tabular}[c]{@{}l@{}}Longest path components that appear\\ in both file paths in relative order but\\ not necessarily contiguous\end{tabular} & \begin{tabular}[c]{@{}l@{}}f1 =“apps/CtsVerifier/src/com/android/cts/verifier/\\ sensors/MagnetometerTestActivity.java”\\ f2 =“tests/tests/hardware/src/android/hardware/cts/\\ SensorTest.java”\\ LCSubseq(f1, f2) = length({[}src, android, cts{]}) = 3\end{tabular} \\ \hline
\end{tabular}
 \label{tab:string_fun}
\end{table}
In this model, the key step is to calculate the file path similarity, with file path $\mathbb{f}_a$ and $\mathbb{f}_b$ computation as follows:
%%%%%%%%%%%%%%%%%%%%%%%%
\begin{equation}
FilePathSimilarity_{LCx}(\mathbb{f}_a, \mathbb{f}_b) = \frac{LCx(\mathbb{f}_a, \mathbb{f}_b)}{\text{max(Length($\mathbb{f}_a$), Length($\mathbb{f}_b$))}}
\end{equation}
%%%%%%%%%%%%%%%%%%%%%%%
where  $LCx(\mathbb{f}_a, \mathbb{f}_b)$ function is a parameter specifying how to compare file path components $\mathbb{f}_a$ and $\mathbb{f}_b$.
%To compare file path components \textcolor{red}{of} $f_n$ and $f_i$,
%similar to we use 
Table \ref{tab:string_fun} presents the definitions and example calculation for four file path comparison techniques.
The four comparison techniques, i.e. Longest Common Prefix (LCP), Longest Common Suffix (LCS), Longest Common Substring (LCSubstr), and Longest Common Subsequence (LCSubseq)~\cite{Gusfield_1997}, are used in the $LCx$ function. 
The comparison function value is normalized by the maximum length of $\mathbb{f}_a$ and $\mathbb{f}_b$, i.e., the number of file path components.

\begin{algorithm}
\footnotesize
    \SetKwInOut{Input}{Input}
    \SetKwInOut{Output}{Output}
    \SetKwInOut{Method}{Method}
    \Input{\noindent A new patch ($\mathbb{p}_n$); A set of patches ($Patches$); A file path comparison function $LCx$ }
    \Output{A list of candidate patches}
    $\mathbb{F}_n \leftarrow Files(\mathbb{p}_n)$\;
    \For{$\mathbb{p}_i \in Patches$}{
        $\mathbb{F}_i \leftarrow Files(\mathbb{p}_i)$\; 
        $SimilaritySum \leftarrow 0$\;
        \For{$\mathbb{f}_n \in \mathbb{F}_n$}{
            \For{$\mathbb{f}_i \in \mathbb{F}_i$}{
                $SimilaritySum \leftarrow SimilaritySum + FilePathSimilarity_{LCx}(\mathbb{f}_n, \mathbb{f}_i)$\;
            }
        }
        $\mathbb{sim}_i \leftarrow \frac{1}{|\mathbb{F}_n| \cdot |\mathbb{F}_i|} SimilaritySum $\;%\revise{explain the goal \textcolor{red}{of} this in the text}
    }
    \Return A patch list in $Patches$ in the descending order of $\mathbb{sim}_i$;
  \caption{Detect linkage based on file location similarity}
  \label{Al2}
\end{algorithm}

Algorithm \ref{Al2} describes the procedure used to calculate file location similarity. 
Inputs include a new patch $\mathbb{p}_n$, a set of patches $Patches$, and a file path comparison function $LCx$.
\textit{Files($\mathbb{p})$} represents extracting a modified file set in a patch $\mathbb{p}$ (line 1 and 3).
Then for each patch $\mathbb{p}_i$ in the $Patches$, the file location similarity score between $\mathbb{p}_n$ and $\mathbb{p}_i$ is computed using the $FilePathSimilarity_{LCx}$ function (lines 4-9). 
Finally, ($\mathbb{sim}_i$) measures the average value of the file path similarity for every file path in $\mathbb{p}_n$ and $\mathbb{p}_i$ (line 10).
After the patches are sorted based on their similarity scores, the algorithm returns a list of candidate patches in descending similarity order (line 12).

\paragraph{Combination Technique - }
Prior work~\cite{Kittler_1998} successfully shows that the model performance can be improved when combining individual techniques.
Similar to \citet{SANER2015_Pick}, our algorithm is based on the  Borda count method~\cite{borda} for scoring the ranks.
The Borda count is a voting technique that simply combines the recommendation lists based on the rank.

\begin{itemize}
\item {\uline{\textit{Combined file location model (fl).}}}
First, we combine the four string comparison techniques (i.e., LCP, LCS, LCSubstr, and LCSubseq) used in the file location model.
For each patch candidate $\mathbb{p}_{k}$, we assign scores based on the rank of $\mathbb{p}_{k}$ in each recommendation list generated from four comparison techniques, with the candidate with the highest ranks receiving the highest scores.
For example, if a recommendation list of $\mathbb{C}_{LCP}$ votes a patch candidate $\mathbb{p}_{1}$ as the first rank and the number of total candidates are $\mathbb{S}$, then this patch candidate will get the score of $\mathbb{S}$.
The candidate $\mathbb{p}_{10}$ will get a score of $\mathbb{S}$ - 10.
Given a set of recommendation lists $\mathbb{C}$  $\in$ \{$\mathbb{C}_{LCP}$, $\mathbb{C}_{LCS}$, $\mathbb{C}_{LCSubstr}$, $\mathbb{C}_{LCSubseq}$\}, the score for a patch candidate $\mathbb{p}_{k}$ is defined as follows:
\begin{equation}
Combination(\mathbb{p}_{k})=\sum_{ \mathbb{C}_{i} \in \mathbb{C}}\mathbb{S}_{i} - rank (\mathbb{p}_{k} | \mathbb{C}_{i})
\label{Combination}
\end{equation}
where $\mathbb{S}_{i}$ is the total number of the recommended patch candidates, rank ($\mathbb{p}_{k}$ | $\mathbb{C}_{i}$) represents the rank of patch candidate $\mathbb{p}_{k}$ in $\mathbb{C}_{i}$.
The patch linkage recommendation is a list of patch candidates that are ranked according to their scores.
To resolve tie-breakers, we reorder candidate patches whose Borda scores are same based on their created time, bubbling up recent patches to the top.
\item {\uline{\textit{Combined File location and Textual Contents.}}}
Second, we combine our candidate lists from the textual content model (tc) and the combined file location model (fl).
Following Equation~\ref{Combination}, we assign scores based on the rank of $\mathbb{p}_{k}$ in each recommendation list from tc and fl ($\mathbb{C}$  $\in$ \{$\mathbb{C}_{tc}$,$\mathbb{C}_{fl}$\}) and rank the patch candidates according to their Borda scores.
Like the combined file location algorithm, we use the patch created time to reorder candidate patches that return the same scores.
\end{itemize}

\begin{table}[pos = b]
\centering
\caption{Dataset used in experiment based on time intervals. Time intervals are divided into 2 days, 7 days, 14 days, and 30 days.}
\label{model_dataset}
\resizebox{\textwidth}{!}{
\begin{tabular}{l|rrr|rrrr|}
\cline{2-8}
                                                 & \multicolumn{3}{c|}{Ground-truth from RQ1} & \multicolumn{4}{c|}{Experiment Dataset}                                   \\ \hline
\multicolumn{1}{|l|}{Porject}                    & Interval (Days) & \#Patches & \#File Paths & \#Avg. Patches & \#Total Patches & \#Avg. File Paths & \#Total File Paths \\
\multicolumn{1}{|l|}{\multirow{4}{*}{Qt}}        & 2               & 40        & 198          & 174            & 6,953           & 1,868             & 74,714             \\
\multicolumn{1}{|l|}{}                           & 7               & 74        & 455          & 528            & 39,033          & 5,948             & 440,115            \\
\multicolumn{1}{|l|}{}                           & 14              & 96        & 528          & 1,018          & 97,723          & 10,718            & 1,028,905          \\
\multicolumn{1}{|l|}{}                           & 30              & 119       & 632          & 2,169          & 258,070         & 23,193            & 2,760,000          \\ \hline
\multicolumn{1}{|l|}{\multirow{4}{*}{OpenStack}} & 2               & 87        & 259          & 605            & 52,637          & 1,755             & 152,644            \\
\multicolumn{1}{|l|}{}                           & 7               & 147       & 689          & 1,659          & 243,791         & 5,199             & 764,209            \\
\multicolumn{1}{|l|}{}                           & 14              & 196       & 811          & 3,398          & 666,040         & 10,681            & 2,093,308          \\
\multicolumn{1}{|l|}{}                           & 30              & 233       & 1,055        & 7,267          & 1693,125        & 23,143            & 5,392,245          \\ \hline
\multicolumn{1}{|l|}{\multirow{4}{*}{AOSP}}      & 2               & 62        & 708          & 117            & 7,259           & 2,493             & 154,539            \\
\multicolumn{1}{|l|}{}                           & 7               & 106       & 971          & 327            & 34,629          & 8,755             & 928,019            \\
\multicolumn{1}{|l|}{}                           & 14              & 137       & 2,759        & 654            & 89,574          & 17,017            & 2,331,345          \\
\multicolumn{1}{|l|}{}                           & 30              & 170       & 2,997        & 1,393          & 236,714         & 31,659            & 5,381,939          \\ \hline
\end{tabular}}
\end{table}

\subsection{Data Preparation}
Table \ref{model_dataset} shows our dataset that is adopted in the model evaluation experiment, using the same three projects from the exploratory study.
To construct a more realistic experiment setting, we now use time intervals.
For the ground-truth, we collect patches that are created in the same time interval (i.e., 2, 7, 14, and 30 days.) with the ground-truth patch as our experiment dataset (i.e., document set for the textual similarity analysis).
Note that within a patch pair, we always treat the patches whose created time are later as our ground truth.
Our ground truth includes patch features (i.e., textual content and file location).
Our assumption is that a closer time interval should have a higher textual and file location similarity for that patch linkage.
For instance, the OpenStack ground truth includes 87 labeled patches (2 days), 147 labeled patches (7 days), 196 labeled patches (14 days), and 233 labeled patches (30 days).

\subsection{Evaluation Metrics}
To evaluate our approach for patch linkage detection, we use recommendation metrics that are commonly used in software engineering domains~\cite{ASE2012_Nguyen, stackover_2016,SANER2015_Pick}.
Since each patch only allows one target patch, other evaluation metrics (i.e., Mean Average Precision) are not suitable for this study.
The metrics are defined below:
\begin{itemize}

\item \uline{\textit{Recall@k}} calculates the relevant items proportion found in the candidate list.
A high recall means that an algorithm returned more relevant results.
%%%%%%%
\begin{subequations}
\begin{equation}
\textit{$Recall@k_{all}$}=\frac{|\mathbb{D}|}{|\mathbb{G}|}
\label{Recall}
\end{equation}
\begin{equation}
\textit{$Recall@k_{type}$}=\frac{|\mathbb{D}_{type}|}{|\mathbb{G}_{type}|}
\label{Accuracy}
\end{equation}
\end{subequations}
%%%%%%%%%%%%

The Equation \ref{Recall} defines how the Recall@k is calculated for all patch linkage types.
In this formula, $\mathbb{D}$ refers to a set of ground truth (patches) whose linked patches are retrieved in the candidate list, while $\mathbb{G}$ refers to a set of ground truth (patches) used in the experiment. 
Equation \ref{Accuracy} computes the Recall@k for specific linkage type.
$\mathbb{D}_{type}$ is a set of ground truth (patches) labeled with one linkage type whose linked patches are retrieved in the candidate list, while $\mathbb{G}_{type}$ refers to a set of ground truth (patches) labeled with one linkage type used in the experiment.
Inspired by the  previous study \cite{review_icse13,expert_09}, we set k to range from 1 to 10.

\item \uline{\textit{$Precision@k_{type}$}} calculates the ratio of correctly predicted positive observations to the total predicted positive observations.
A high precision relates to the low false-positive rate.

\begin{equation}
\textit{$Precision@k_{type}$}=\frac{|\mathbb{D}_{type}|}{|\mathbb{D}|}
\label{Precision}
\end{equation}

Equation \ref{Precision} defines our precision calculation for specific linkage type, where $\mathbb{D}_{type}$ refers to a set of ground truth (patches) labeled with one linkage type whose linked patches are retrieved in the candidate list, while
$\mathbb{D}$ refers to a set of ground truth (patches) whose linked patches are retrieved in the candidate list.

\item \uline{\textit{Mean Reciprocal Rank (MRR@k)}} calculates an average of the reciprocal ranks for correctly detected patches in a candidate list.
A high MRR score indicates that the first true positive is being returned closer to the top list.

\begin{equation}
\textit{MRR@k}=\frac{1}{|\mathbb{G}|}\sum_{\mathbb{g} \in \mathbb{G}}^{}\frac{{1}}{rank(candidates(\mathbb{g}))}
\label{MRR}
\end{equation}

Equation \ref{MRR} explains the MRR@k calculation. Given a set of ground truth $\mathbb{G}$ (patches) used in the experiment, the $rank (candidates(\mathbb{g}))$ refers to the rank position value of a ground truth $\mathbb{g}$ (a patch)  whose linked patch is retrieved in the candidate list.
If there is no correctly retrieved patch in the candidate list, the value of  $\frac{1}{rank(candidates(\mathbb{g}))}$ will be 0.
\end{itemize}

\subsection{{\textbf{Results for RQ2}}}
\begin{figure}[pos = t]
    \centering
    \includegraphics[width=0.9\linewidth]{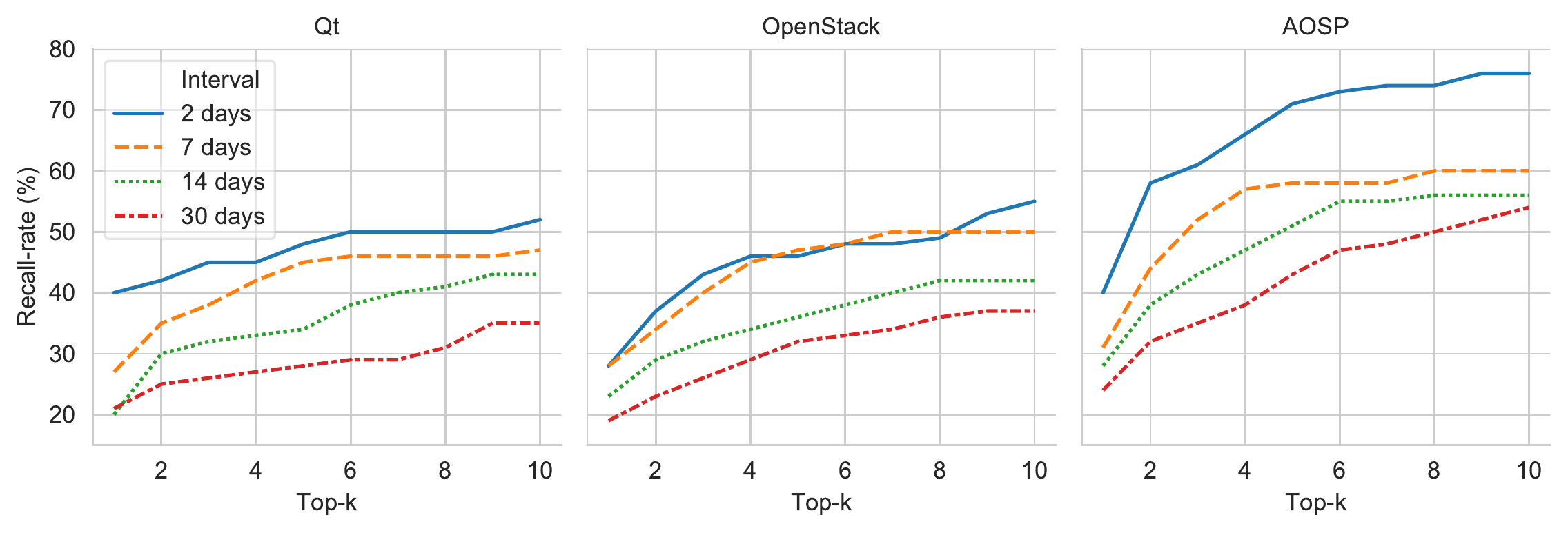}
    \caption{$Recall@k_{all}$ for the detection based on textual content (the concatenation of title and description text in a patch). The results suggest that the recall rates for the textual content model decrease when the time intervals get larger.}
    \label{Rq1_fig}
\end{figure}

\begin{table}[pos = b]
\footnotesize
\centering
\caption{Evaluation results ($Recall@k_{type}$ and $MRR@10$) for the textual content model. The recall rates for detecting the Alternative Solution linkage range from 34\% to 69\%, 34\% to 80\%, and 31\% to 82\% for Qt, OpenStack, and AOSP.}
\resizebox{\textwidth}{!}{
\begin{tabular}{ll|rrrr|rrrr|rrrr|r|}
\cline{3-15}
                                                &                 & \multicolumn{4}{c|}{Alternative Solution Recall}    & \multicolumn{4}{c|}{Broader Context Recall}         & \multicolumn{4}{c|}{Dependency Recall}              & \multicolumn{1}{c|}{MRR}    \\ \hline
\multicolumn{1}{|l}{Project}                    & Interval (Days) & Top-1 & Top-3 & Top-5 & \multicolumn{1}{l|}{Top-10} & Top-1 & Top-3 & Top-5 & \multicolumn{1}{l|}{Top-10} & Top-1 & Top-3 & Top-5 & \multicolumn{1}{l|}{Top-10} & \multicolumn{1}{l|}{Top-10} \\ \hline
\multicolumn{1}{|l}{\multirow{4}{*}{Qt}}        & 2               & 58\%  & 63\%  & 69\%  & 69\%                        & 34\%  & 45\%  & 45\%  & 56\%                        & 17\%  & 17\%  & 17\%  & 25\%                        & 43\%                        \\
\multicolumn{1}{|l}{}                           & 7               & 45\%  & 62\%  & 65\%  & 68\%                        & 20\%  & 25\%  & 40\%  & 45\%                        & 9\%   & 18\%  & 22\%  & 22\%                        & 34\%                        \\
\multicolumn{1}{|l}{}                           & 14              & 34\%  & 54\%  & 54\%  & 61\%                        & 16\%  & 24\%  & 32\%  & 44\%                        & 3\%   & 10\%  & 10\%  & 16\%                        & 27\%                        \\
\multicolumn{1}{|l}{}                           & 30              & 35\%  & 45\%  & 47\%  & 55\%                        & 19\%  & 23\%  & 23\%  & 32\%                        & 5\%   & 5\%   & 8\%   & 13\%                        & 25\%                        \\ \hline
\multicolumn{1}{|l}{\multirow{4}{*}{OpenStack}} & 2               & 54\%  & 67\%  & 70\%  & 80\%                        & 13\%  & 38\%  & 41\%  & 47\%                        & 17\%  & 21\%  & 25\%  & 38\%                        & 36\%                        \\
\multicolumn{1}{|l}{}                           & 7               & 56\%  & 70\%  & 73\%  & 79\%                        & 13\%  & 28\%  & 40\%  & 44\%                        & 18\%  & 25\%  & 29\%  & 29\%                        & 36\%                        \\
\multicolumn{1}{|l}{}                           & 14              & 46\%  & 62\%  & 62\%  & 63\%                        & 15\%  & 20\%  & 29\%  & 40\%                        & 14\%  & 19\%  & 22\%  & 26\%                        & 29\%                        \\
\multicolumn{1}{|l}{}                           & 30              & 34\%  & 49\%  & 56\%  & 60\%                        & 15\%  & 18\%  & 22\%  & 30\%                        & 12\%  & 14\%  & 20\%  & 24\%                        & 24\%                        \\ \hline
\multicolumn{1}{|l}{\multirow{4}{*}{AOSP}}      & 2               & 43\%  & 68\%  & 75\%  & 82\%                        & 30\%  & 50\%  & 65\%  & 65\%                        & 50\%  & 64\%  & 71\%  & 79\%                        & 53\%                        \\
\multicolumn{1}{|l}{}                           & 7               & 42\%  & 61\%  & 70\%  & 74\%                        & 14\%  & 42\%  & 45\%  & 47\%                        & 38\%  & 50\%  & 54\%  & 54\%                        & 42\%                        \\
\multicolumn{1}{|l}{}                           & 14              & 43\%  & 56\%  & 67\%  & 71\%                        & 10\%  & 30\%  & 37\%  & 40\%                        & 26\%  & 36\%  & 39\%  & 48\%                        & 38\%                        \\
\multicolumn{1}{|l}{}                           & 30              & 31\%  & 48\%  & 57\%  & 64\%                        & 13\%  & 22\%  & 31\%  & 46\%                        & 22\%  & 27\%  & 33\%  & 43\%                        & 32\%                        \\ \hline
\end{tabular}}
\label{RQ2_result}
\end{table}

\uline{\textit{Results for RQ2.1}} --
To evaluate the performance of the textual content model, we compute the $Recall@k_{all}$ for each project, the $Recall@k_{type}$ for each patch linkage type, and the $Mean Reciprocal Rank$ ($MRR$) scores.
Figure~\ref{Rq1_fig} and Table \ref{RQ2_result} show the evaluation results of the textual content model.
We summarize two main findings from this table.

The recall rates for the textual content model decrease when the time intervals get larger.
Figure~\ref{Rq1_fig} shows the recall rates of the textual content model covering all patch linkage types from Top-1 to Top-10 for Qt, OpenStack, and AOSP.
We observe that when the time interval is set as 2 days, the recall rates can range from 40\% to 53\%,  28\% to 55\%, and 40\% to 76\% for these projects.
While when the time intervals are set as 7, 14, 30 days, the recall rates keep increasing compared to the interval of 2 days.
For instance, in the interval of 30 days, the recall rates vary from 21\% to 35\%, 19\% to 38\%, and 24\% to 54\% for the three projects.
These results indicate that the selection of the time interval affects the textual content model performance, as more submitted patches could potentially increase the complexity of the textual corpus.

The detection of the Alternative Solution linkage outperforms the other patch linkage types with relatively high recall rates.
Table \ref{RQ2_result} reports the recall rates for the Top-1, Top-3, Top-5, and Top-10 of three studied linkage types.
We observe that the recall rates of the Alternative Solution linkages can range from 34\% to 69\%, 34\% to 80\%, and 31\% to 82\% for Qt, OpenStack, and AOSP.
On the other hand, the recall rates of the Broader Context linkage vary from 16\% to 56\%, 13\% to 47\%, and 10\% to 65\%.
For the Dependency linkage, the recall rates range from 3\% to 25\%, 12\% to 38\%, and 22\% to 79\%.
These results suggest that the linked patches with an Alternative Solution linkage share more similar textual contents than the other two linkage types.

\uline{\textit{Results for RQ2.2}} --
We use the combination technique as described in Section 4.1 to combine recommendation lists from four string comparison techniques (i.e., LCP, LCS, LCSubstr, and LCSubseq).
To evaluate the performance of the file location model, similarly, we compute the $Recall@k_{all}$ for each project, the $Recall@k_{type}$ for each patch linkage type, and the $Mean Reciprocal Rank$ ($MRR$) scores.
Figure~\ref{Rq3_result} and Table \ref{RQ3} show the evaluation results of the file location model.
We now discuss our findings below.

\begin{figure}[pos = t]
    \centering
    \includegraphics[width=0.9\linewidth]{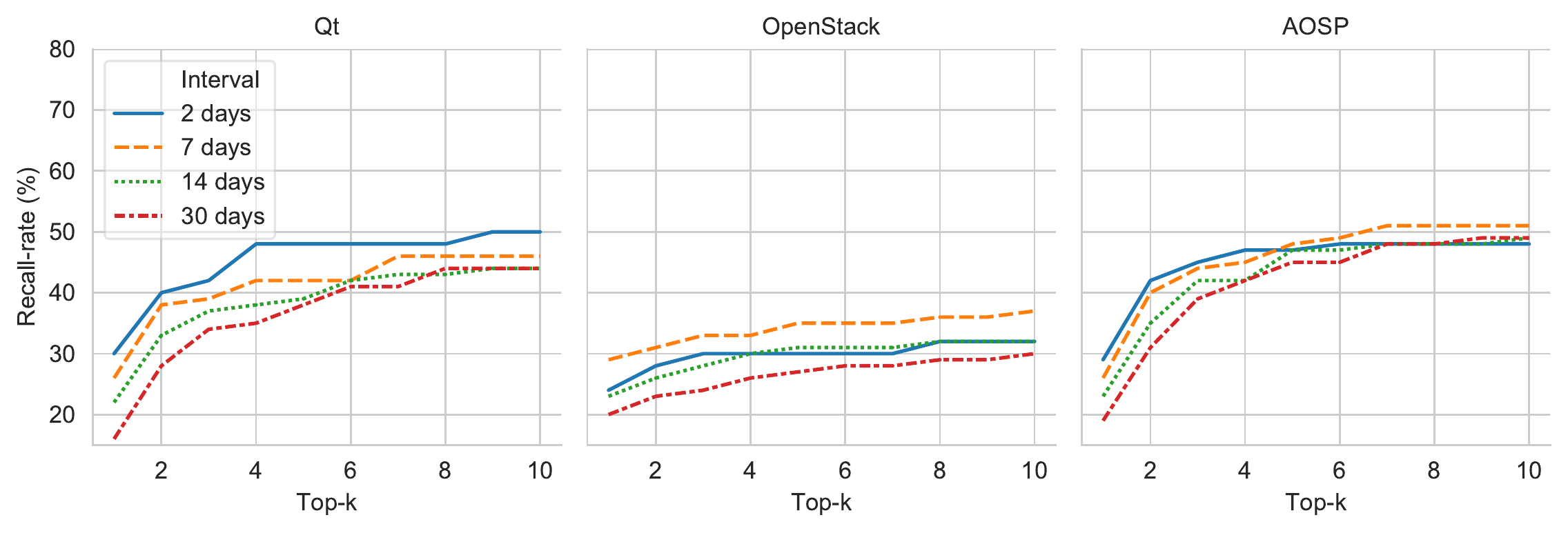}
    \caption{$Recall@k_{all}$ for the detection based on file location (a set of file paths that the patch modifies). The results show that the file location as a patch feature overall does not perform as well as the textual content feature with relatively lower recall rates (16\%--50\% for Qt, 19\%--37\% for OpenStack, and 19\%--51\% for AOSP).}
    \label{Rq3_result}
\end{figure}

\begin{table}[pos = b]
\centering
\caption{Evaluation results ($Recall@k_{type}$ and $MRR@10$) for the file location model. The recall rates for detecting the Alternative Solution linkage range from 23\% to 69\%, 34\% to 60\%, and 26\% to 72\% for Qt, OpenStack, and AOSP.}
\label{RQ3}
\resizebox{\textwidth}{!}{
\begin{tabular}{ll|rrrr|rrrr|rrrr|r|}
\cline{3-15}
                                                &          & \multicolumn{4}{c|}{Alternative Solution Recall}    & \multicolumn{4}{c|}{Broader Context Recall}         & \multicolumn{4}{c|}{Dependency Recall}              & \multicolumn{1}{c|}{MRR}    \\ \hline
\multicolumn{1}{|l}{Project}                    & Interval (days) & Top-1 & Top-3 & Top-5 & \multicolumn{1}{l|}{Top-10} & Top-1 & Top-3 & Top-5 & \multicolumn{1}{l|}{Top-10} & Top-1 & Top-3 & Top-5 & \multicolumn{1}{l|}{Top-10} & \multicolumn{1}{l|}{Top-10} \\ \hline
\multicolumn{1}{|l}{\multirow{4}{*}{Qt}}        & 2        & 37\%  & 53\%  & 63\%  & 69\%                        & 22\%  & 45\%  & 45\%  & 45\%                        & 25\%  & 25\%  & 25\%  & 25\%                        & 50\%                       \\
\multicolumn{1}{|l}{}                           & 7        & 29\%  & 42\%  & 45\%  & 55\%                        & 20\%  & 45\%  & 50\%  & 50\%                        & 26\%  & 31\%  & 31\%  & 31\%                        & 34\%                        \\
\multicolumn{1}{|l}{}                           & 14       & 24\%  & 46\%  & 49\%  & 56\%                        & 24\%  & 44\%  & 48\%  & 48\%                        & 16\%  & 19\%  & 26\%  & 26\%                        & 30\%                        \\
\multicolumn{1}{|l}{}                           & 30       & 23\%  & 41\%  & 47\%  & 57\%                        & 13\%  & 39\%  & 42\%  & 45\%                        & 10\%  & 21\%  & 21\%  & 26\%                        & 26\%                        \\ \hline
\multicolumn{1}{|l}{\multirow{4}{*}{OpenStack}} & 2        & 34\%  & 47\%  & 47\%  & 50\%                        & 24\%  & 24\%  & 24\%  & 27\%                        & 13\%  & 17\%  & 17\%  & 17\%                        & 27\%                        \\
\multicolumn{1}{|l}{}                           & 7        & 47\%  & 53\%  & 55\%  & 60\%                        & 22\%  & 27\%  & 29\%  & 29\%                        & 18\%  & 18\%  & 20\%  & 22\%                        & 31\%                        \\
\multicolumn{1}{|l}{}                           & 14       & 40\%  & 51\%  & 53\%  & 56\%                        & 19\%  & 23\%  & 28\%  & 28\%                        & 12\%  & 14\%  & 15\%  & 15\%                        & 30\%                        \\
\multicolumn{1}{|l}{}                           & 30       & 35\%  & 40\%  & 46\%  & 52\%                        & 17\%  & 20\%  & 23\%  & 25\%                        & 9\%   & 11\%  & 14\%  & 16\%                        & 23\%                        \\ \hline
\multicolumn{1}{|l}{\multirow{4}{*}{AOSP}}      & 2        & 43\%  & 64\%  & 68\%  & 72\%                        & 20\%  & 30\%  & 30\%  & 30\%                        & 14\%  & 29\%  & 29\%  & 29\%                        & 37\%                        \\
\multicolumn{1}{|l}{}                           & 7        & 37\%  & 59\%  & 63\%  & 68\%                        & 17\%  & 36\%  & 36\%  & 36\%                        & 21\%  & 29\%  & 38\%  & 42\%                        & 36\%                        \\
\multicolumn{1}{|l}{}                           & 14       & 32\%  & 57\%  & 59\%  & 61\%                        & 9\%   & 30\%  & 37\%  & 37\%                        & 23\%  & 26\%  & 32\%  & 39\%                        & 33\%                        \\
\multicolumn{1}{|l}{}                           & 30       & 26\%  & 49\%  & 55\%  & 60\%                        & 9\%   & 29\%  & 36\%  & 40\%                        & 19\%  & 33\%  & 35\%  & 41\%                        & 30\%                        \\ \hline
\end{tabular}}
\end{table}

The detection using file location feature overall can not perform as well as the model using the textual content feature.
Figure~\ref{Rq3_result} shows the recall rates of the file location model from Top-1 to Top-10 for Qt, OpenStack, and AOSP.
As we can see, the overall performance is visually lower than the model using textual content (except for the cases in the intervals of 14 days and 30 days for Qt).
For example, when the interval is set as 2 days, the recall rates range from 30\% to 50\%, 24\% to 32\%, and 29\% to 48\% for the three projects.
Table~\ref{RQ3} shows the recall rates for each patch linkage type using file location.
We observe that the Alternative Solution linkage still outperforms the other linkage types, i.e., the Recall@10 rates of 69\%, 50\%, and 72\% in the interval of 2 days for three projects.
Such result indicates that the linked patches with an Alternative Solution linkage are more likely to modify similar file locations.

\begin{figure}[pos = hbt!]
    \centering
    \includegraphics[width=.88\linewidth]{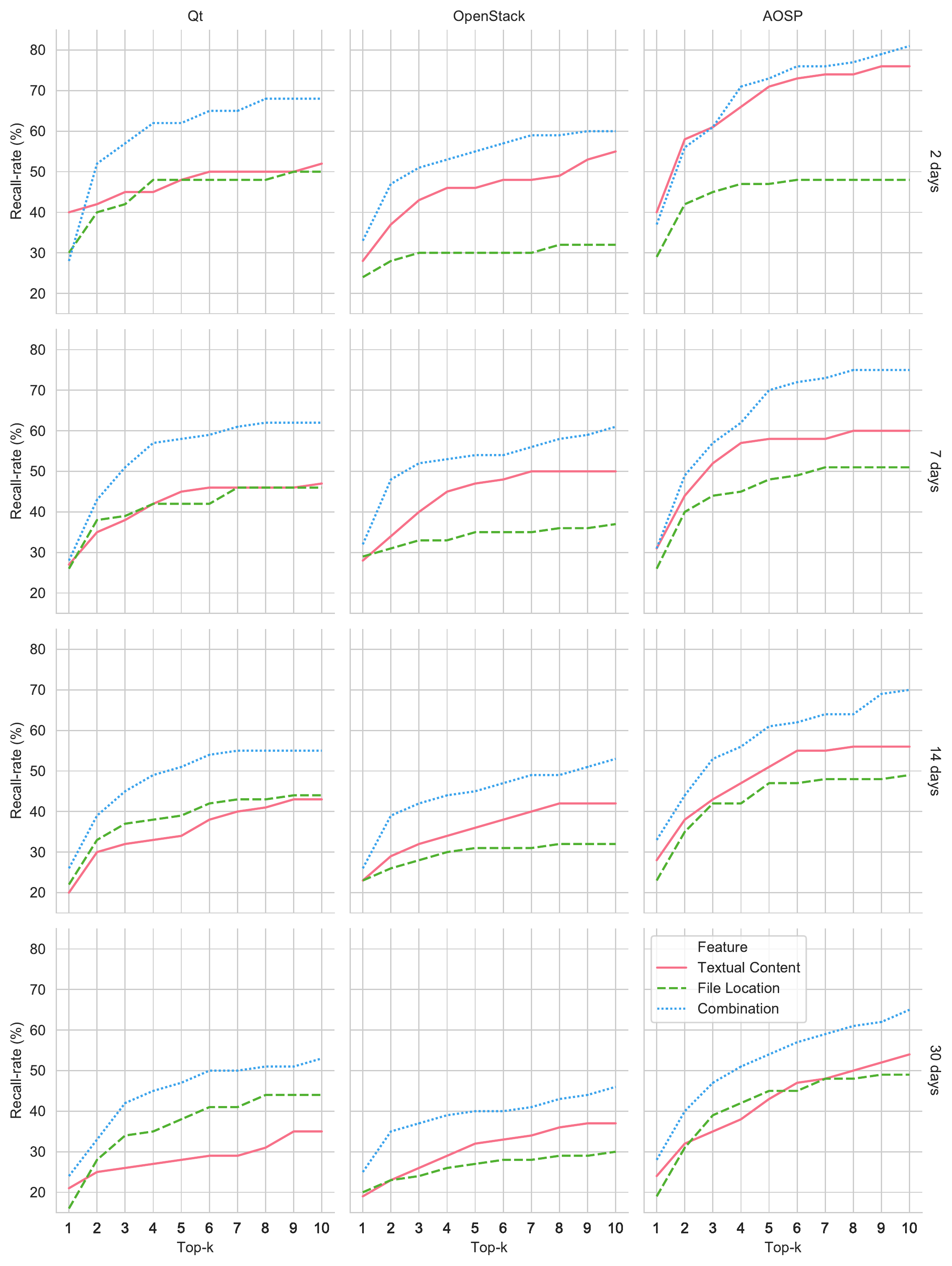}
    \caption{$Recall@k_{all}$ for the detection based on file location and textual content for studied projects. The results suggest that when we combine two features together, the detection model can achieve higher recall rates in high ranking positions, compared to the separate models.}
    \label{combination}
\end{figure}

\uline{\textit{Results for RQ2.3}} --
We use the combination technique as described in Section 4.1 to combine two features, based on two Top-10 candidate lists from the separate models (textual content model and file location model).
To evaluate the performance of the model combined with two features, we compute the $Recall@k_{all}$ for the three projects, the $Recall@k_{type}$ for linkage types, and the $Mean Reciprocal Rank$ ($MRR$) scores.
In addition, we would like to investigate the  $Precision@k_{type}$  of the Alternative Solution linkage, since the separate models were able to detect the Alternative Solution linkage best.
Figure~\ref{combination}, Table 8, and Table 9 show the evaluation results related to the feature combination model.
We now discuss two main findings below.

\begin{table}[pos=b]
\centering
\caption{Evaluation results ($Recall@k_{type}$ and $MRR@10$) for the feature combination model. The higher $MRR@10$ scores show that the model can detect more patch linkages in higher ranks (i.e., 33\%--44\% for Qt, 33\%--43\% for OpenStack, and 40\%--53\% for AOSP).}
\label{RQ4}
\resizebox{\textwidth}{!}{
\begin{tabular}{ll|rrrr|rrrr|rrrr|r|}
\cline{3-15}
                                                &          & \multicolumn{4}{c|}{Alternative Solution Recall}    & \multicolumn{4}{c|}{Broader Context Recall}         & \multicolumn{4}{c|}{Dependency Recall}              & \multicolumn{1}{c|}{MRR}    \\ \hline
\multicolumn{1}{|l}{Project}                    & Interval (days) & Top-1 & Top-3 & Top-5 & \multicolumn{1}{l|}{Top-10} & Top-1 & Top-3 & Top-5 & \multicolumn{1}{l|}{Top-10} & Top-1 & Top-3 & Top-5 & \multicolumn{1}{l|}{Top-10} & \multicolumn{1}{l|}{Top-10} \\ \hline
\multicolumn{1}{|l}{\multirow{4}{*}{Qt}}        & 2        & 32\%  & 74\%  & 84\%  & 95\%                        & 33\%  & 67\%  & 67\%  & 67\%                        & 17\%  & 25\%  & 25\%  & 25\%                        & 44\%                       \\
\multicolumn{1}{|l}{}                           & 7        & 35\%  & 68\%  & 78\%  & 81\%                        & 30\%  & 45\%  & 55\%  & 60\%                        & 17\%  & 35\%  & 35\%  & 39\%                        & 41\%                        \\
\multicolumn{1}{|l}{}                           & 14       & 37\%  & 59\%  & 66\%  & 76\%                        & 28\%  & 52\%  & 56\%  & 56\%                        & 10\%  & 23\%  & 26\%  & 26\%                        & 36\%                       \\
\multicolumn{1}{|l}{}                           & 30       & 39\%  & 55\%  & 63\%  & 76\%                        & 23\%  & 48\%  & 52\%  & 55\%                        & 5\%   & 21\%  & 23\%  & 23\%                        & 33\%                        \\ \hline
\multicolumn{1}{|l}{\multirow{4}{*}{OpenStack}} & 2        & 50\%  & 77\%  & 84\%  & 84\%                        & 21\%  & 39\%  & 46\%  & 55\%                        & 29\%  & 34\%  & 34\%  & 38\%                        & 43\%                        \\
\multicolumn{1}{|l}{}                           & 7        & 60\%  & 83\%  & 83\%  & 85\%                        & 20\%  & 42\%  & 47\%  & 60\%                        & 18\%  & 34\%  & 34\%  & 36\%                        & 43\%                        \\
\multicolumn{1}{|l}{}                           & 14       & 47\%  & 70\%  & 74\%  & 79\%                        & 21\%  & 34\%  & 39\%  & 50\%                        & 12\%  & 25\%  & 27\%  & 31\%                        & 35\%                        \\
\multicolumn{1}{|l}{}                           & 30       & 39\%  & 59\%  & 64\%  & 71\%                        & 22\%  & 31\%  & 34\%  & 41\%                        & 14\%  & 23\%  & 24\%  & 29\%                        & 33\%                        \\ \hline
\multicolumn{1}{|l}{\multirow{4}{*}{AOSP}}      & 2        & 50\%  & 72\%  & 79\%  & 89\%                        & 25\%  & 45\%  & 65\%  & 75\%                        & 29\%  & 64\%  & 71\%  & 71\%                        & 53\%                        \\
\multicolumn{1}{|l}{}                           & 7        & 44\%  & 70\%  & 87\%  & 94\%                        & 17\%  & 47\%  & 58\%  & 58\%                        & 29\%  & 46\%  & 54\%  & 63\%                        & 46\%                        \\
\multicolumn{1}{|l}{}                           & 14       & 49\%  & 68\%  & 79\%  & 90\%                        & 14\%  & 40\%  & 49\%  & 56\%                        & 26\%  & 39\%  & 42\%  & 48\%                        & 44\%                        \\
\multicolumn{1}{|l}{}                           & 30       & 41\%  & 64\%  & 68\%  & 77\%                        & 16\%  & 35\%  & 44\%  & 56\%                        & 24\%  & 35\%  & 46\%  & 54\%                        & 40\%                       \\ \hline
\end{tabular}}
\end{table}

\begin{table}[pos = b]
\centering
\caption{Evaluation results ($Precision@k_{type}$) for the Alternative Solution linkage. The results show that precision is relatively higher in the separate models than in the feature combination models (i.e., 60\%--74\% and 43\%--67\% in the textual content model for Qt and OpenStack; 56\%--67\% in the file location model for AOSP).}
\label{RQ4_precision}
\resizebox{\textwidth}{!}{
\begin{tabular}{ll|rrrr|rrrr|rrrr|}
\cline{3-14}
                                                &                 & \multicolumn{4}{c|}{Textual Content}                & \multicolumn{4}{c|}{File Location}                  & \multicolumn{4}{c|}{Combination}                    \\ \hline
\multicolumn{1}{|l}{Project}                    & Interval (Days) & Top-1 & Top-3 & Top-5 & \multicolumn{1}{l|}{Top-10} & Top-1 & Top-3 & Top-5 & \multicolumn{1}{l|}{Top-10} & Top-1 & Top-3 & Top-5 & \multicolumn{1}{l|}{Top-10} \\ \hline
\multicolumn{1}{|l}{\multirow{4}{*}{Qt}}        & 2               & 69\%  & 67\%  & 68\%  & 62\%                        & 58\%  & 59\%  & 68\%  & 70\%                        & 55\%  & 61\%  & 64\%  & 67\%                        \\
\multicolumn{1}{|l}{}                           & 7               & 70\%  & 68\%  & 63\%  & 60\%                        & 47\%  & 45\%  & 45\%  & 50\%                        & 52\%  & 55\%  & 56\%  & 54\%                        \\
\multicolumn{1}{|l}{}                           & 14              & 74\%  & 71\%  & 67\%  & 61\%                        & 47\%  & 53\%  & 53\%  & 54\%                        & 60\%  & 55\%  & 55\%  & 59\%                        \\
\multicolumn{1}{|l}{}                           & 30              & 68\%  & 71\%  & 72\%  & 64\%                        & 58\%  & 50\%  & 53\%  & 54\%                        & 68\%  & 54\%  & 55\%  & 59\%                        \\ \hline
\multicolumn{1}{|l}{\multirow{4}{*}{OpenStack}} & 2               & 67\%  & 54\%  & 53\%  & 50\%                        & 48\%  & 54\%  & 54\%  & 54\%                        & 52\%  & 52\%  & 52\%  & 48\%                        \\
\multicolumn{1}{|l}{}                           & 7               & 64\%  & 56\%  & 49\%  & 50\%                        & 52\%  & 52\%  & 51\%  & 52\%                        & 60\%  & 51\%  & 49\%  & 45\%                        \\
\multicolumn{1}{|l}{}                           & 14              & 57\%  & 56\%  & 50\%  & 43\%                        & 51\%  & 51\%  & 48\%  & 48\%                        & 53\%  & 49\%  & 47\%  & 44\%                        \\
\multicolumn{1}{|l}{}                           & 30              & 53\%  & 57\%  & 53\%  & 49\%                        & 54\%  & 53\%  & 52\%  & 52\%                        & 48\%  & 49\%  & 49\%  & 47\%                        \\ \hline
\multicolumn{1}{|l}{\multirow{4}{*}{AOSP}}      & 2               & 48\%  & 53\%  & 50\%  & 49\%                        & 67\%  & 64\%  & 66\%  & 67\%                        & 61\%  & 53\%  & 49\%  & 52\%                        \\
\multicolumn{1}{|l}{}                           & 7               & 58\%  & 51\%  & 53\%  & 53\%                        & 61\%  & 57\%  & 57\%  & 58\%                        & 61\%  & 53\%  & 54\%  & 54\%                        \\
\multicolumn{1}{|l}{}                           & 14              & 69\%  & 59\%  & 60\%  & 59\%                        & 65\%  & 63\%  & 59\%  & 58\%                        & 69\%  & 60\%  & 60\%  & 59\%                        \\
\multicolumn{1}{|l}{}                           & 30              & 63\%  & 63\%  & 61\%  & 55\%                        & 63\%  & 58\%  & 57\%  & 56\%                        & 64\%  & 61\%  & 56\%  & 54\%                        \\ \hline
\end{tabular}}
\end{table}

The model combined with two features performs better than two separate models using a single feature.
Figure~\ref{Combination} shows the combination model performance for three projects at different time intervals.
From the interval of 2 days to the interval of 30 days, the feature combination model performs with a recall rate, ranging from 24\% to 68\% for Qt, 25\% to 61\% for OpenStack, and 28\% to 81\% for AOSP.
Moreover, Table 8 records the $MRR@10$ scores for each studied project.
Compared to the $MRR@10$ scores in separate models, we find that the $MRR@10$ scores as well increase in the feature combination model.
For instance, in the OpenStack project, the $MRR@10$ scores are between 33\% and 43\% while the scores are between 24\% and 36\%, 23\% and 31\% in the textual content model and file location model.
The higher MRR scores indicate that the feature combination model can detect more patch linkages with higher ranks.

The detection of the Alternative Solution linkage is promising with high recall rates and possible precision rates.
Table \ref{RQ4} shows the recall rates of three linkage types detected at different intervals.
We observe that with the combination of two features, the recall rates of the Alternative Solution linkage increase to 76\%--95\%, 71\%--85\%, and 77\%--94\% in the Top-10 for Qt, OpenStack, and Android, which is much higher than the other two patch linkages (i.e., 23\%--39\%, 29\%--38\%, and 48\%--71\% for the Dependency linkage).
Table \ref{RQ4_precision} specifically calculates the $Precision@k_{type}$ for the Alternative Solution linkage.
The table shows that overall the precision does not increase in the combination model compared to those separate models.
For example, for Qt and OpenStack projects, the textual content model generally performs better with the precision rates being from 60\% to 74\% and 43\% to 67\%.
On the other hand, for AOSP project, the file location model achieves higher precision rates, ranging from 56\% to 67\%. 

\begin{tcolorbox}
\textbf{Answering RQ2:} 
Results show that combining two features (i.e., textual content and file location) performs better than two separate models.
In experiments that span four time intervals (i.e., 2, 7, 14, and 30 days), the model performs with promising recall rates (i.e., 24\%--68\% for Qt, 25\%--61\% for OpenStack, and 28\%--81\% for AOSP).
The Alternative Solution linkage detection is also promising with relatively high recall rates (i.e., 74\%--95\% for Qt, 71\%--87\% for OpenStack, and 77\%--94\% for AOSP in the Top-10).
Reasonable Alternative Solution linkage detection also means that the precision rates are feasible,  with 60\%--74\% for Qt, 43\%--67\% for OpenStack in the textual content model, and  56\%--67\% for AOSP in the file location model.
\end{tcolorbox}

\section{Discussion}
\label{sec:implications}
We now discuss four insights from our empirical experiments and the interpretation of the results.
\begin{enumerate}
    \item \uline{\textit{Latency of patch linkage notification.}}
    The exploratory results provide evidence that there is latency until the review team is notified with the patch linkage.
   \citet{dataset_pull} similarly recognized this detection latency in the case of duplicate pull requests. 
   For researchers, we believe that this study can be used to motivate the need for early detection of potential patch linkages, which could be in the form of automatic tool support.
   Furthermore, we are unsure if the latency is due to team awareness or that the review is not yet fully understood by the team that is reviewing the patch.
   For the review team, maybe part of the workflow should include the search for similar patches that could be linked, especially those patches that introduce alternative solutions. 
   Potential future work would be studying reasons for the latency and its impact on the review process.
   A possible example includes latency caused by waiting for a reviewer.
   
\item \uline{\textit{Team awareness enhancement}}. 
Active awareness can improve teams' trust, relationships, and efficiency ~\cite{awareness_01} as review discussions particularly serve as an important mechanism for coordination and collaboration between team members.
\citet{Bacchelli_ICSE2013} reported that practitioners at Micros{of}t view team awareness as an important motivation for conducting code review.
According to the results in our exploratory study, we notice that the patch with an Alternative solution linkage tends to finish the review process quicker after the linkage has been established. 
For researchers, understanding the role that linkages play in team awareness might be a potential future research avenue.
For the practitioners, we suggest that improving the awareness between the patches may also increase the likelihood that the linked patches will be identified.
This can be done by making sure there are overlaps in the review teams.

\item \uline{\textit{Detection improvements.}}
Results show that using more than one feature (i.e., textual content and file location) can improve performance.
We also find that the textual information in a patch is complex and very difficult to standardize English.
As such, using the typical information retrieval methods might not be ideal.
To improve the textual content model, developers could be encouraged to increase the natural language or generate a project specific corpus.
For researchers, other more sophisticated textual similarity could be calculated.
Furthermore, looking at the similarity of the patch contents themselves (i.e., source code)  might be interesting for improvement of the model.
For developers, our research shows the potential for related patches, which is already adopted in practice.
For example, the Gerrit OpenStack web interface is able to show  (i) related patches and (ii) same topic patches for a patch under review.

\item \uline{\textit{Detection in Practice.}}
The linkage detection is promising in a realistic setting, especially for the Alternative Solution linkage.
Our results in Table~\ref{RQ4} and Table~\ref{RQ4_precision} from RQ2 show that the detection of the Alternative Solution linkage can reach the recall rates of 74\%--95\%, 71\%--87\%, and 77\%--94\% in the Top-10 for Qt, OpenStack, and AOSP.
Although the precision rates are possible (60\%--74\%, Qt; 43\%--67\%, OpenStack; 56\%--67\%, AOSP), we acknowledge that our model still has room for improvement.
One possible clue is to set the threshold for the textual similarity, as the similarity shared by different linkage types could be different.

At the same time, we would like to investigate whether there is a difference between the realistic experiment and the sample-based experiment.
To do so, following the technique employed by previous work, we apply the 20\% and 80\% ratio to construct the sample-based datasets.
In detail, 20\% are ground truth patches with labeled linkage types, while 80\% are randomly selected patches with no linkages. 
Table~\ref{tab:sample_base} shows the model evaluation results using sample-based datasets.
Two observations are summarized from the table.
The first observation is that the file location models perform better than the textual content models.
In contrast, in our time interval based experiment, the results suggest that the textual content models outperform.
One potential reason leading to the difference could be that the file location is not as flexible as the textual content in the reality.
The one same file location can be edited by more than one developer with different purposes.
The second observation is that in all, the experiment based on the time intervals performs better than the one using the sample-based datasets.
This observation indicates that patches that are closer in specific time intervals are more likely to share similar textual contents and file locations than patches that are relatively longer time apart.
Inspired by the existing results from the time interval based experiment, the immediate future work is to find out whether or not the implementation of our proposed models would be useful by studying real users.

\begin{table}[pos = t]
\centering
\caption{Evaluation results using sample-based datasets. The statistics show that in the sample-based evaluation, the file location model performs better than the textual content model.}
\label{tab:sample_base}
\resizebox{\textwidth}{!}{
\begin{tabular}{ll|rrrr|rrrr|rrrr|rrrr|}
\cline{3-18}
                                                &                 & \multicolumn{4}{c|}{All}       & \multicolumn{4}{c|}{Alternative Solution} & \multicolumn{4}{c|}{Broader Context} & \multicolumn{4}{c|}{Dependency} \\ \hline
\multicolumn{1}{|l}{Project}                    & Feature         & Top-1 & Top-3 & Top-5 & Top-10 & Top-1    & Top-3    & Top-5    & Top-10   & Top-1   & Top-3   & Top-5  & Top-10  & Top-1  & Top-3 & Top-5 & Top-10 \\ \hline
\multicolumn{1}{|l}{\multirow{2}{*}{Qt}}        & Textual Content & 24\%  & 32\%  & 36\%  & 43\%   & 39\%     & 45\%     & 53\%     & 61\%     & 14\%    & 23\%    & 27\%   & 35\%    & 15\%   & 23\%  & 23\%  & 27\%   \\
\multicolumn{1}{|l}{}                           & File Location   & 31\%  & 44\%  & 48\%  & 56\%   & 44\%     & 56\%     & 61\%     & 71\%     & 23\%    & 36\%    & 39\%   & 45\%    & 22\%   & 37\%  & 42\%  & 47\%   \\
\multicolumn{1}{|l}{\multirow{2}{*}{OpenStack}} & Textual Content & 24\%  & 32\%  & 37\%  & 44\%   & 45\%     & 58\%     & 64\%     & 68\%     & 16\%    & 24\%    & 28\%   & 34\%    & 13\%   & 19\%  & 25\%  & 25\%   \\
\multicolumn{1}{|l}{}                           & File Location   & 29\%  & 39\%  & 44\%  & 52\%   & 53\%     & 61\%     & 66\%     & 75\%     & 20\%    & 31\%    & 35\%   & 44\%    & 19\%   & 29\%  & 35\%  & 42\%   \\
\multicolumn{1}{|l}{\multirow{2}{*}{Android}}   & Textual Content & 29\%  & 41\%  & 47\%  & 56\%   & 35\%     & 52\%     & 58\%     & 64\%     & 18\%    & 30\%    & 36\%   & 49\%    & 32\%   & 37\%  & 40\%  & 47\%   \\
\multicolumn{1}{|l}{}                           & File Location   & 40\%  & 51\%  & 57\%  & 65\%   & 54\%     & 65\%     & 72\%     & 79\%     & 28\%    & 38\%    & 46\%   & 53\%    & 26\%   & 40\%  & 40\%  & 53\%   \\ \hline
\end{tabular}}
\end{table}

\end{enumerate}

\section{Threats to Validity}
\label{sec:threats}
In this section, we describe the threats to the validity of both exploratory study and linkage detection study.

\uline{\textit{Construct validity.}}
We summarize the construct validity into two parts according to our exploratory study and patch linkage detection study.
In the exploratory study, the threat exists in our approach to recover patch linkages.
We use regular expressions to identify the patch linkages in the form of Changeid, Review \#, and hyperlinks.
However, it is prone to generate false positives which means these formats cannot assure the target is a patch, i.e., Review \# could be bug id.
To mitigate the risk of such false positives, we conducted a careful manual check as described in Section 3.2 (DP1) Patch linkage recovery and filtering.
With regards to the patch linkage detection, one threat is the time interval selection in our experiment dataset, as we cannot cover all patch linkages using the current time intervals.
However, we believe that such an experiment setting is closed to reality and could provide insight into whether or not the time interval has influence on the linkage detection.
Another threat related to the detection experiment is patch feature selection as we do not include other features such as source code.
Our assumption is that the file location provides some heuristic of the source code.
We will list this as an immediate future study to improve our models.

\uline{\textit{Internal validity.}}
In our exploratory study, a potential threat exists in our qualitative method to manually classify types of patch linkages.
Patch linkages may be mis-coded due to the subjective nature of understanding the coding schema by the co-authors.
To mitigate this threat, we took a systematic approach to first test our comprehension with 30 samples using Kappa agreement scores by three separate individuals. 
Only until the Kappa score reaches more than 0.8 (nearly perfect), we were able to complete the rest of the sample dataset.
Another threat is related to the tool selection for our study, as they may change the results of the study.
In our study, we rely on various tools such as gensim for the tf-idf calculation and the Mann-Whitney U test for the statistical significance testing.
We are however confident, as these tools have been widely used in other studies.

\uline{\textit{External validity.}}
External validity is related to the result generalization.
Our empirical findings are based on three open source projects using the Gerrit code review tool, i.e., Qt, OpenStack, and AOSP.
However, it is unknown whether our results can be generalized to the other tool-based code review such as pull-based code review.
For those smaller commercial projects or projects that adopt the commit-then-review style, unawareness of patch linkage could not be an issue.
Thus, the patch linkage issue might only affect large open source projects or projects that adopt the review-then-commit style.
Due to the popular rise of tools (i.e., Gerrit and Pull-Requests) that support the review-then-commit style, 
we believe that the problem still does affect many software development teams.
In order to encourage future replication studies, our replication package is available at \textcolor{black}{\url{https://github.com/dong-w/Replication-Patch-Linkage}}.

\section{Related Work}
\label{sec:relatework}
In this section, we will discuss the related work, in terms of code review models, duplicate pull request detection, and duplicate bug report detection.

\uline{\textit{Code review models - } }
Code review is used to examine the changes made by other developers, to find potential defects and improve project quality since the 1970s \cite{Fagan_1976}. 
In an earlier time, it was well known as code inspection.
Many tools were developed to support the formal process of code inspection~\cite{icicle,csi,Scrutiny,tool_97icse}.
Over the last decade, code review relying on modern review tools has been widely adopted by many open source projects~\cite{rigby_2012}, i.e., Gerrit tool in OpenStack, pull request in Github projects, CodeFlow in Microsoft~\cite{Bosu_useful}.
There are two review styles: review-then-commit (RTC) and commit-then-review (CTR).
In the pioneering work, \citet{rigby_apache_icse} empirically examined the comparison between these two review techniques.
For the style of CTR, projects allow trusted developers to commit contributions before they are reviewed.
In contrast to CTR, RTC is a technique where a review is made before committing it to the original code~\cite{rigby_rtc_tosem}.
For instance, the Android project adopts RTC style using the Gerrit tool to provide a discussion platform for the review process before the patch is merged into the codebase~\cite{rigby_2013_MSR}.
Another example is that GitHub projects apply pull-based development to conduct the code review~\cite{G_14,icse16_pr}.
The study of open source projects that use either the Gerrit tools or GitHub pull requests has been extensively studied \cite{FSE2013_Rigby,SANER2015_Pick,ICSME18_pull}.
Our proposed solution is not applicable to all projects.
For instance, \citet{PROFES_survey_2017} reported that about 50\% of commercial teams use CTR review style.
However, with the growth of large open source projects, it is possible that teams are not able to be aware of the related patches in RTC.
\textit{Our work} focuses on the risk when team awareness is lacking, which is only applicable for large teams or reviewing models that are RTC. 
This includes all Gerrit tools and all GitHub Open Source projects, as they are pull-based developments.

\uline{\textit{Duplicate pull request detection -}}
Within code review models, the topic of the duplicate pull request is specifically investigated.
\citet{dataset_pull} created a dataset extracted from 26 open source projects in Github by using a semi-automatic approach.
Their analysis found 21\% of duplicates were identified after a relatively long latency (more than one week).
Additionally, their statistics show that the redundant review efforts were spent on the duplicates (i.e., on average 2.5 reviewers participating in the redundant review discussions and 5.2 review comments are generated before the duplicate relation is identified.
To address the duplicate pull request detection, by now there are two threads of work: one is using information retrieval and the other one is using classification.
For the information retrieval thread, \citet{pull_request17} used text information of pull request to detect duplicates on three popular projects hosted in GitHub.
The evaluation shows that about 55.3\%--71.0\% of the duplicates can be found when we use the combination of title similarity and description similarity.
For the classification thread, \citet{ren_saner2019} calculated the similarity of nine pull request features where title and description are both included, and then they adopted a machine learning algorithm to aggregate the nine features.
The result shows that the proposed classifiers achieve 57--83\% precision for detecting duplicate code changes from the maintainer's perspective, which outperforms the Li et al.’s work~\cite{pull_request17}.
Recently, \citet{xinxia_pull} integrated the time feature to the nine features proposed by Ren et al. and the experimental results show that it can substantially improve the performance of Ren et al.’s work by 14.36\% and 11.93\% in terms of F1-score@1 and F1-score@5, respectively.
\textit{Our work} is different from the duplicate pull request detection, as our focus is not only limited to the duplication, but also consider other patch linkages (i.e., alternative solution, broader context, and dependency).

\uline{\textit{Duplicate bug report detection -}}
Although the duplicate pull request detection is not widely studied, there has been much work on investigating the detection of other duplicate artifacts in SE domains, such as duplicate bug reports in the issue tracking system.
Bug reports provide textual description that involves the
natural language bug description reported by developers, i.e.,
title and description. 
Natural language information and information retrieval (IR) techniques are widely used to calculate the similarity scores between a given data and the retrieved data.
For instance, \citet{Runeson07_ICSE} took natural language text of bug reports and performed standard tokenization, stemming, and stop word removal.
Their results recognize the importance of using textual information with two-thirds of the duplicates can be found through NLP technologies.
\citet{harm_icsme2008} studied the importance of duplicate bug reports and found that the additional information provided by duplicates helps to resolve bugs quicker.
At the same time, their classification model achieved an average precision and recall of 68\% and 60\%.
Researchers also found that in addition to natural language information, other extra report information can improve duplicate retrieval.
\citet{icse08wang} combined natural language and execution information to detect the duplicate reports.
The approach they proposed can detect more duplication compared to only using natural language information alone (i.e., 67\%--73\% of duplication can be detected).
\citet{ASE_11} proposed REP to improve the accuracy of duplicate bug retrieval.
Not only text but also other available information such as product, component, and priority are fully utilized in REP.

\section{Conclusion}
\label{sec:conclusion}
In this work, we perform an exploratory study to understand the impact of patch linkage on the review process and conduct experiments to evaluate automatic patch linkage detection based on realistic time intervals.
Our study provides evidence that latency exists in the notification of a patch linkage, and confirms that patch linkage detection is promising, with likely improvements if the practice of posting linkages becomes more prevalent. 
This study provides many open avenues for future work, which includes (i) studying the role that linkages play in a review team awareness, (ii) improvement of the detection using alternative approaches with additional features (i.e., source code), (iii) improving our feature metrics, and (iv) studying real users through the implementation of our proposed models. 
From our discussions, this paper lays the groundwork for future research on how to increase patch linkage awareness, which may facilitate a more efficient review.

\section*{Acknowledgement}
This work is supported by Japanese Society for the Promotion of Science (JSPS) KAKENHI Grant Numbers JP18H04094, JP20K19774 and JP20H05706.

\bibliographystyle{cas-model2-names}

\bibliography{cas-refs}

\end{document}